\documentclass{LMCS}

\usepackage{amssymb,amsfonts,amsmath,eepic,epic,latexsym}
\usepackage{graphicx,epsfig,hyperref}

\newcommand{\act}{{\it Act}}
\newcommand{\ball}[2]{{\it B}_{#1}({#2})}
\newcommand{\bx}[1]{{[{#1}]}}
\newcommand{\ce}[1]{{\bf K}({#1})}
\newcommand{\compact}[2]{{\mathcal C}[{#1},{#2}]}
\newcommand{\defeq}{=}
\newcommand{\denote}[2]{{[\!\mid {#1}\mid \!]_{#2}}}
\newcommand{\denoteD}[1]{{\{\!\mid {#1}\mid \!\}}}
\newcommand{\denotemode}[2]{{[\!\mid {#1}\mid \!]^{{\it #2}}}}
\newcommand{\dia}[1]{{\langle {#1}\rangle }}
\newcommand{\down}[2]{{\downarrow \!\! _{#2}\!{#1}}}
\newcommand{\e}{\{\}}
\newcommand{\embed}[1]{{\langle\!\mid\! {#1}\!\mid\!\rangle}}
\newcommand{\false}{{\it f\!f}}
\newcommand{\formula}[3]{{\psi ^{{#1},\, {#2}}_{#3}}}

\newcommand{\idom}{{\mathbb I}}
\renewcommand{\inf}{\mbox{{\it inf}\,}}

\newcommand{\lawson}[1]{{\lambda _{#1}}}
\newcommand{\lowpart}[2]{{#1}^{\it a}_{#2}}
\newcommand{\lpd}[1]{{\mathcal L} [{#1}]}
\newcommand{\meu}[1]{{{\it M}({#1})}}
\newcommand{\modeup}[2]{{#1} ^{\it {#2}}}
\newcommand{\move}[1]{}
\newcommand{\mpa}{{\it MPA}}
\newcommand{\mpd}[1]{{\mathcal M} [{#1}]}
\newcommand{\mts}[1]{{\mathcal {#1}}}
\newcommand{\mub}[1]{{\it mub}({#1})}
\newcommand{\mubi}[2]{{\it mub}^{#1}({#2})}
\newcommand{\prt}[3]{\displaystyle\frac{\,#1\,}{\,#2\,}{\,\scriptstyle #3}}
\newcommand{\ps}[1]{{\mathbb P}({#1})}
\newcommand{\refine}[1]{{\prec}_{#1}}
\newcommand{\reject}[1]{}
\newcommand{\Real}{{\mathbb R}}
\newcommand{\s}{\subseteq}
\newcommand{\scott}[1]{{\sigma _{#1}}}
\renewcommand{\sup}{{\it sup}\,}
\newcommand{\true}{{\it t\!t}}
\newcommand{\ub}[1]{{\it ub}({#1})}
\newcommand{\univ}{{\mathbb D}}
\newcommand{\univx}{{\mathbb X}}
\newcommand{\upp}[2]{{\uparrow \!\! _{#2}\!{#1}}}
\newcommand{\uppart}[2]{{#1}^{\it c}_{#2}}

\def\doi{1 (1:1) 2005}
\lmcsheading%
{\doi}
{28}
{}
{}
{Sep.~\phantom{0}1, 2004}
{Jan.~26, 2005}
{}

\begin{document}

\title[Labelled transition systems as a Stone space]{Labelled transition systems as a Stone space}

\author[M.\ Huth]{Michael Huth}   
\address{Department of Computing, Imperial College London, South 
Kensington campus,\hfill\break London SW7 2AZ, United Kingdom} 
\email{M.Huth@doc.imperial.ac.uk}  



\begin{abstract}
  \noindent A fully abstract and universal domain model for modal
  transition systems and refinement, developed in \cite{hjs04}, is
  shown to be a maximal-points space model for the bisimulation
  quotient of labelled transition systems over a finite set of
  events. In this domain model we prove that this quotient is a Stone
  space whose compact, zero-dimensional, and ultra-metrizable
  Hausdorff topology measures the degree of bisimilarity and that
  image-finite labelled transition systems are dense. Using this
  compactness we show that the set of labelled transition systems that
  refine a modal transition system, its ``set of implementations,'' is
  compact and derive a compactness theorem for Hennessy-Milner logic
  on such implementation sets. These results extend to systems that
  also have partially specified state propositions, unify existing
  denotational, operational, and metric semantics on partial
  processes, render robust consistency measures for modal transition
  systems, and yield an abstract interpretation of compact sets of
  labelled transition systems as Scott-closed sets of modal transition
  systems.
\end{abstract}

\maketitle

\section{Introduction}\label{section:i} Labelled transition
systems are a fundamental modelling formalism in many areas of
computer science and one often needs to compare two or more such
systems in applications. For example, in doing state compression
prior to model checking one wants to ensure that the compressed
system yields the same model checks as the uncompressed one.
Similarly, if one system is a specification and another one its
implementation, then program correctness can be established by
proving these systems to be equivalent. By the same token, if two
systems are not equivalent, one may want to know to what degree
this is so, e.g.\ in a risk analysis of a safety-critical system.

This paper chooses bisimulation as the notion of equivalence of
labelled transition systems.\footnote{Weak bisimulation
\cite{milner89} is bisimulation on a modified transition relation
and we don't consider fairness in this paper.} Bisimulation is an
established, sufficiently fine-grained notion of equivalence
between labelled transition systems \cite{milner89} so any
approximative notions, e.g.\ testing \cite{nicola84}, have
bisimulation as a well accepted point of reference. Since
quantitative aspects ought to be invariant under bisimulation, we
stipulate that the quotient of all labelled transition systems
with respect to bisimulation is the right conceptual space for
reasoning about and comparing quantitative aspects of labelled
transition systems.

If two labelled transition systems are not bisimilar, one may
require a quantitative measure of such differences and such a
measure has many applications. We mention security protocols
\cite{ryan01}, where one system is the specification and the other
is an implementation and where we may wish to quantify illicit
information flow \cite{dipierrojcs02} or the effort needed to
expose implementation flaws; modal specifications
\cite{larsen89b}, where a specification captures a possibly
infinite set of mutually non-bisimilar labelled transition
systems; and requirements engineering \cite{gause89}, where each
system may be the modal specification of a particular viewpoint
and \emph{consistency measures on modal specifications} are
sought.

One principal aim of this paper is to unify several strands of
established work in one integrated framework: metric semantics of
processes \`a la Bakker \& Zucker \cite{BZ_acm82}; use of
Hennessy-Milner logic, domain theory and transition systems \`a la
Abramsky \cite{abramsky91}; means of under-specifying and refining
processes \`a la Larsen \& Thomsen \cite{larsen88}; and
representations of classical topological spaces as maximal-point
spaces of domains \`a la Lawson \cite{lawson97}. To that end, we
use a domain $\univ$, defined in \cite{hjs04} and shown to be a
universal model for finitely-branching modal transition systems
and fully abstract for their refinement in loc.\ cit.

Specifically, we discover that the metric induced by the Lawson
topology on $\univ$ is a generalization of the one in
\cite{BZ_acm82} to modal transition systems; that the subspace of
maximal elements of $\univ$ is a Stone space with respect to the
Lawson (or Scott) topology; and that this Stone space is an
isomorphic representation of the quotient of all labelled
transition systems modulo bisimulation, so the topology and metric
carry over to that quotient. Since a Stone space has a
\emph{complete} ultra-metric, our model has labelled transition
systems that are not image-finite, allowing the modelling of
continuous state spaces, but all labelled transition systems can
be approximated by image-finite ones to any degree of precision.

The compactness of this quotient space then makes it possible to
study the topological structure of sets of implementations for
modal transition systems, the second principal aim of this paper.
In particular, our topological analysis shows that 3-valued model
checking \cite{bruns99,bruns2000} reasons about compact sets of
labelled transition systems, namely the set of all 2-valued
refinements of a given 3-valued system. We propose two measures, a
pessimistic and an optimistic one, for how close any refining
labelled transition systems of two such 3-valued systems can be.
Using compactness, we prove that the optimistic measure is zero
iff the two 3-valued systems in question have a common refinement.

Our concepts and results are also \emph{robust under a change of
representation}, e.g.\ in moving from event-based to state-based
systems or those that combine state and event information. It
would be of interest to see whether results similar to the ones of
this paper are obtainable for systems that explicitly represent
time, probability (e.g.\ as done in \cite{radha00,dipierrojcs02})
or other quantitative information.

\medskip {\bf\sl Outline of this paper:} In Section~\ref{section:domain} we
review modal transition systems, their refinement, and a fully
abstract domain model for these notions.
Section~\ref{section:stone} establishes the central result of this
paper, showing that the maximal-points space of the fully abstract
domain of Section~\ref{section:domain} is a Stone space and the
quotient of all labelled transition systems with respect to
bisimulation. In Section~\ref{section:applications} we give three
applications of the compactness of this maximal-points space: a
compactness theorem for Hennessy-Milner logic on compact sets of
implementations, an abstract interpretation of compact sets of
implementations as Scott-closed sets of modal transition systems,
and a robust consistency measure for modal transition systems.
Section~\ref{section:related} states related work, and
Section~\ref{section:conclude} concludes.

\section{Domain of modal transition systems}\label{section:domain}
Modal transition systems \cite{larsen88} are defined like labelled
transition systems, except that transitions come in two modes that
specify whether such transitions must or may be implemented. A
refinement relation between modal transition systems therefore
associates to a modal transition system those refining labelled
transition systems in which all implementation choices have been
resolved. In this section we formalize these notions and present
the domain of \cite{hjs04} as a faithful mathematical model of the
model-checking framework of modal transition systems.

\subsection{Mixed transition systems and refinement}

We define Larsen \& Thomsen's modal transition systems
\cite{larsen88}, their refinement and other key concepts formally
and present the domain $\univ$ which is a fully abstract model of
such systems and their refinement \cite{hjs04}. Our results are
shown within that domain. In this paper, let $(\alpha,\beta,\dots
\in )\act$ be a fixed finite set of events and $(w, w',\dots
\in)\act ^*$ the set of finite words over $\act$ with $\epsilon$
denoting the word of length zero. The labelled transition systems
considered here have events from $\act$ only. The structural
properties of our domain model require that we also define Dams'
more general notion of mixed transition systems
\cite{dams96,damsGerthGrumberg97}.

A modal transition system $M$ has two transition relations
$R^a,R^c \s\Sigma\times\act\times\Sigma$ on a set of states
$\Sigma$. The sets $\modeup Ra$ and
$\Sigma\times\act\times\Sigma\setminus \modeup Rc$ specify
\emph{contractual promises or expectations} about the reactive
capacity and incapacity of implementations, respectively. These
guarantees are to be understood with respect to the refinement of
states. We write ``{\it a}'' in $R^a$ to denote {\bf a}sserted
behavior and ``{\it c}'' in $R^c$ to denote {\bf c}onsistent
behavior and use these annotations in judgments $\models ^a$ and
$\models ^c$ below with the same meaning.

\begin{exa}
In Figure~\ref{fig:ref} we see a contractual guarantee that any
state refining ${\rm Drinks}$ cannot have a transition labelled
with ${\rm newPint}$ to a state refining ${\rm Talks}$ as the
triple $({\rm Drinks}, {\rm newPint}, {\rm Talks})$ is not in
$\modeup Rc$. There is a contractual guarantee that any state
refining ${\rm Waits}$ has a $\modeup Ra$-transition labelled with
${\rm newPint}$ to all states that refine ${\rm Drinks}$ or ${\rm
Talks}$.
\end{exa}

\begin{defi}\hfill 
\begin{enumerate}
\item \begin{itemize} \item A \emph{mixed transition system}
\cite{dams96,damsGerthGrumberg97} is a triple \(M = (\Sigma,
\modeup Ra,\modeup Rc)\) such that, for every \emph{mode} \({\it
m}\in \{ {\it a}, {\it c}\}\), the pair \((\Sigma,\modeup Rm)\) is
a \emph{labelled transition system}, i.e.\ $\modeup
Rm\s\Sigma\times\act\times\Sigma$.

\item If \(\modeup Ra\s \modeup Rc\), then $M$ is a \emph{modal}
transition system \cite{larsen88}.

\item We call $M$ \emph{image-\-finite} iff for all $s\in \Sigma$,
$\alpha\in\act$, and ${\it m}\in \{{\it a}, {\it c}\}$ the set
\(\{s'\in\Sigma\mid (s,\alpha,s')\in \modeup Rm\}\) is finite.

\item A mixed transition system $M$ with a designated initial
state $i$ is \emph{pointed}, written $(M,i)$.

\item We call elements of $\modeup Ra$ \emph{must-transitions} and
elements of $\modeup Rc\setminus \modeup Ra$
\emph{may-transitions}.
\end{itemize}

\item Let \(M = (\Sigma,\modeup Ra,\modeup Rc)\) be a mixed
transition system.

\begin{itemize}
\item A relation \(Q\s \Sigma\times\Sigma\) is a \emph{refinement
within \(M\)} \cite{larsen88,dams96} iff \((s,t)\in Q\) implies,    
for all $\alpha\in\act$,

\begin{enumerate}
\item if \((s,\alpha,s')\in \modeup Ra\), there exists some
\((t,\alpha,t')\in\modeup Ra\) such that  \((s',t')\in Q\); and
\item if \((t,\alpha,t')\in \modeup Rc\), there exists some
\((s,\alpha,s')\in\modeup Rc\) such that \((s',t')\in Q\).

\end{enumerate}

\item We write \(s\,\refine{M}\,t\) or \(s\,\refine{}\,t\) if
there is some refinement $Q$ with $(s,t)\in Q$. In that case, $t$
\emph{refines} (\emph{is abstracted by}) $s$.

\item States $s$ and $t$ are \emph{refinement-equivalent} iff
($s\refine{} t$ and $t\refine{} s$).

\item Let $(M,i)\refine{} (N,j)$ mean that $j$ refines $i$ in the
mixed transition system that is the disjoint union of $M$ and $N$;
$(M,i)$ and $(N,j)$ are refinement-equivalent iff $i$ and $j$ are
refinement-equivalent in that union.

\item The \emph{implementations} of $(M,i)$ are those pointed
modal transition systems without may-transitions that refine
$(M,i)$.

\end{itemize}

\end{enumerate}
\end{defi}

\noindent As the union $\refine M$ of all refinements within $M$
is also a refinement within $M$, $\refine M$ is the greatest
refinement relation within $M$. Please note that we use the
relational inverse of the $Q$ in \cite{larsen88,dams96,hjs04}, as
done in \cite{gj02}, so our $(M,i)\refine {}(N,j)$ is written as
$(N,j)\refine {} (M,i)$ in \cite{hjs04}. Larsen \& Thomsen's modal
transition systems and their refinement \cite{larsen88} are
partial versions of labelled transition systems and bisimulation
\cite{milner89}. A modal transition system represents those
labelled transition systems that refine it, the implementations of
$M$. This representation is sound, for if a modal transition
system $M$ refines a modal transition system $N$, all labelled
transition systems that refine $M$ also refine $N$ as $\refine{}$
is transitive.

\begin{exa}\hfill 
\begin{enumerate}
\item Figures~\ref{fig:ref} and~\ref{fig:reftwo} depict modal
transition systems, where dashed and solid lines depict
may-transitions and must-transitions, respectively. The refinement
$Q$ identifies states with the same activity; e.g.\ {\rm Drinks}
with {\rm TomDrinks} and {\rm BobDrinks} etc.

\item The mixed transition system on the left of
Figure~\ref{fig:mixed} is not a modal transition system but is
refinement-equivalent to the modal transition system on the right
of Figure~\ref{fig:mixed}.
\end{enumerate}
\end{exa}

\begin{figure}
\begin{center}
\setlength{\unitlength}{0.00052493in}
\begingroup\makeatletter\ifx\SetFigFont\undefined%
\gdef\SetFigFont#1#2#3#4#5{%
  \reset@font\fontsize{#1}{#2pt}%
  \fontfamily{#3}\fontseries{#4}\fontshape{#5}%
  \selectfont}%
\fi\endgroup%
{\renewcommand{\dashlinestretch}{30}
\begin{picture}(3471,2056)(0,-10)
\dashline{60.000}(957,1129)(2982,1129)
\blacken\path(2862.000,1099.000)(2982.000,1129.000)(2862.000,1159.000)(2898.000,1129.000)(2862.000,1099.000)
\dashline{60.000}(597,1039)(1542,364)
\blacken\path(1426.915,409.337)(1542.000,364.000)(1461.789,458.161)(1473.646,412.824)(1426.915,409.337)
\dashline{60.000}(3297,994)(2037,364)
\blacken\path(2130.915,444.498)(2037.000,364.000)(2157.748,390.833)(2112.132,401.566)(2130.915,444.498)
\dashline{60.000}(3207,1309)(1722,1759)(822,1309)
\blacken\path(915.915,1389.498)(822.000,1309.000)(942.748,1335.833)(897.132,1346.566)(915.915,1389.498)
\dashline{60.000}(462,1309)(12,1579)(372,2029)(597,1354)
\blacken\path(530.592,1458.355)(597.000,1354.000)(587.513,1477.329)(570.437,1433.689)(530.592,1458.355)
\path(1632,274)(597,274)(597,859)
\blacken\path(627.000,739.000)(597.000,859.000)(567.000,739.000)(597.000,775.000)(627.000,739.000)
\path(2172,274)(3297,274)(3297,814)
\blacken\path(3327.000,694.000)(3297.000,814.000)(3267.000,694.000)(3297.000,730.000)(3327.000,694.000)
\put(1677,184){\makebox(0,0)[lb]{\smash{{{\SetFigFont{8}{9.6}{\rmdefault}{\mddefault}{\updefault}Waits}}}}}
\put(372,1084){\makebox(0,0)[lb]{\smash{{{\SetFigFont{8}{9.6}{\rmdefault}{\mddefault}{\updefault}Drinks}}}}}
\put(3072,1084){\makebox(0,0)[lb]{\smash{{{\SetFigFont{8}{9.6}{\rmdefault}{\mddefault}{\updefault}Talks}}}}}
\put(1677,1219){\makebox(0,0)[lb]{\smash{{{\SetFigFont{8}{9.6}{\rmdefault}{\mddefault}{\updefault}talks}}}}}
\put(1092,769){\makebox(0,0)[lb]{\smash{{{\SetFigFont{8}{9.6}{\rmdefault}{\mddefault}{\updefault}orders}}}}}
\put(2217,769){\makebox(0,0)[lb]{\smash{{{\SetFigFont{8}{9.6}{\rmdefault}{\mddefault}{\updefault}orders}}}}}
\put(687,49){\makebox(0,0)[lb]{\smash{{{\SetFigFont{8}{9.6}{\rmdefault}{\mddefault}{\updefault}newPint}}}}}
\put(2352,49){\makebox(0,0)[lb]{\smash{{{\SetFigFont{8}{9.6}{\rmdefault}{\mddefault}{\updefault}newPint}}}}}
\put(2397,1669){\makebox(0,0)[lb]{\smash{{{\SetFigFont{8}{9.6}{\rmdefault}{\mddefault}{\updefault}drinks}}}}}
\put(552,1804){\makebox(0,0)[lb]{\smash{{{\SetFigFont{8}{9.6}{\rmdefault}{\mddefault}{\updefault}drinks}}}}}
\end{picture}
}
\end{center}
\caption{An image-\-finite modal transition system specifying
aspects of ``pub behavior.'' \label{fig:ref}}
\end{figure}
\begin{figure}
\begin{center}
\setlength{\unitlength}{0.00052493in}
\begingroup\makeatletter\ifx\SetFigFont\undefined%
\gdef\SetFigFont#1#2#3#4#5{%
  \reset@font\fontsize{#1}{#2pt}%
  \fontfamily{#3}\fontseries{#4}\fontshape{#5}%
  \selectfont}%
\fi\endgroup%
{\renewcommand{\dashlinestretch}{30}
\begin{picture}(3855,4494)(0,-10)
\dashline{60.000}(777,912)(1722,1992)
\blacken\path(1665.557,1881.936)(1722.000,1992.000)(1620.402,1921.446)(1666.686,1928.784)(1665.557,1881.936)
\dashline{60.000}(732,597)(372,237)(867,12)(867,507)
\blacken\path(897.000,387.000)(867.000,507.000)(837.000,387.000)(867.000,423.000)(897.000,387.000)
\path(3432,3612)(1992,4062)(867,3612)
\blacken\path(967.275,3684.421)(867.000,3612.000)(989.559,3628.713)(944.992,3643.197)(967.275,3684.421)
\path(867,3297)(1767,2262)
\blacken\path(1665.620,2332.867)(1767.000,2262.000)(1710.897,2372.238)(1711.881,2325.387)(1665.620,2332.867)
\path(1677,2127)(642,2127)(642,3207)
\blacken\path(672.000,3087.000)(642.000,3207.000)(612.000,3087.000)(642.000,3123.000)(672.000,3087.000)
\path(552,3612)(12,3972)(642,4467)(642,3792)
\blacken\path(612.000,3912.000)(642.000,3792.000)(672.000,3912.000)(642.000,3876.000)(612.000,3912.000)
\path(3432,3297)(2082,2262)
\blacken\path(2158.980,2358.820)(2082.000,2262.000)(2195.486,2311.204)(2148.663,2313.108)(2158.980,2358.820)
\path(2217,2127)(3567,2127)(3567,3207)
\blacken\path(3597.000,3087.000)(3567.000,3207.000)(3537.000,3087.000)(3567.000,3123.000)(3597.000,3087.000)
\path(1857,1947)(1857,1137)(1182,867)
\blacken\path(1282.275,939.421)(1182.000,867.000)(1304.559,883.713)(1259.992,898.197)(1282.275,939.421)
\path(1992,1947)(1992,1137)(3027,822)
\blacken\path(2903.464,828.239)(3027.000,822.000)(2920.934,885.640)(2946.639,846.458)(2903.464,828.239)
\path(1407,732)(2982,732)
\blacken\path(2862.000,702.000)(2982.000,732.000)(2862.000,762.000)(2898.000,732.000)(2862.000,702.000)
\dashline{60.000}(1362,3432)(3072,3432)
\blacken\path(2952.000,3402.000)(3072.000,3432.000)(2952.000,3462.000)(2988.000,3432.000)(2952.000,3402.000)
\put(1767,2037){\makebox(0,0)[lb]{\smash{{{\SetFigFont{8}{9.6}{\rmdefault}{\mddefault}{\updefault}Waits}}}}}
\put(417,3387){\makebox(0,0)[lb]{\smash{{{\SetFigFont{8}{9.6}{\rmdefault}{\mddefault}{\updefault}BobDrinks}}}}}
\put(417,687){\makebox(0,0)[lb]{\smash{{{\SetFigFont{8}{9.6}{\rmdefault}{\mddefault}{\updefault}TomDrinks}}}}}
\put(3117,3387){\makebox(0,0)[lb]{\smash{{{\SetFigFont{8}{9.6}{\rmdefault}{\mddefault}{\updefault}BobTalks}}}}}
\put(3117,687){\makebox(0,0)[lb]{\smash{{{\SetFigFont{8}{9.6}{\rmdefault}{\mddefault}{\updefault}TomTalks}}}}}
\put(777,4197){\makebox(0,0)[lb]{\smash{{{\SetFigFont{8}{9.6}{\rmdefault}{\mddefault}{\updefault}drinks}}}}}
\put(2532,3972){\makebox(0,0)[lb]{\smash{{{\SetFigFont{8}{9.6}{\rmdefault}{\mddefault}{\updefault}drinks}}}}}
\put(2577,2217){\makebox(0,0)[lb]{\smash{{{\SetFigFont{8}{9.6}{\rmdefault}{\mddefault}{\updefault}newPint}}}}}
\put(867,2217){\makebox(0,0)[lb]{\smash{{{\SetFigFont{8}{9.6}{\rmdefault}{\mddefault}{\updefault}newPint}}}}}
\put(1362,2847){\makebox(0,0)[lb]{\smash{{{\SetFigFont{8}{9.6}{\rmdefault}{\mddefault}{\updefault}orders}}}}}
\put(1767,3522){\makebox(0,0)[lb]{\smash{{{\SetFigFont{8}{9.6}{\rmdefault}{\mddefault}{\updefault}talks}}}}}
\put(642,1452){\makebox(0,0)[lb]{\smash{{{\SetFigFont{8}{9.6}{\rmdefault}{\mddefault}{\updefault}orders}}}}}
\put(1092,102){\makebox(0,0)[lb]{\smash{{{\SetFigFont{8}{9.6}{\rmdefault}{\mddefault}{\updefault}drinks}}}}}
\put(1767,552){\makebox(0,0)[lb]{\smash{{{\SetFigFont{8}{9.6}{\rmdefault}{\mddefault}{\updefault}talks}}}}}
\put(2082,1182){\makebox(0,0)[lb]{\smash{{{\SetFigFont{8}{9.6}{\rmdefault}{\mddefault}{\updefault}newPint}}}}}
\put(2307,2847){\makebox(0,0)[lb]{\smash{{{\SetFigFont{8}{9.6}{\rmdefault}{\mddefault}{\updefault}orders}}}}}
\put(1137,1182){\makebox(0,0)[lb]{\smash{{{\SetFigFont{8}{9.6}{\rmdefault}{\mddefault}{\updefault}newPint}}}}}
\end{picture}
}
\end{center}
\caption{An image-\-finite modal transition system that refines
the one in Figure~\ref{fig:ref}. \label{fig:reftwo}}
\end{figure}
\begin{rem}
We may identify modal transition systems $(\Sigma, R, R)$ with
labelled transition systems $(\Sigma, R)$ and refinement between
such modal transition systems with bisimulation \cite{larsen88}
and will freely move between these two representations of labelled
transition systems and bisimulation subsequently.
\end{rem}

\subsection{The interval domain as an allegory}

Before we present the domain model for refinement of modal
transition systems we use Scott's interval domain \cite{scott72}
as a motivating example that features most of the desirable
properties of our domain model.

\begin{exa}
Figure~\ref{fig:idom} shows the interval domain and its ordering:
$[r,s]\leq [r',s']$ iff ($r\leq r'$ and $s'\leq s$). In that case
we say that $[r',s']$ refines $[r,s]$. \end{exa}
\begin{figure}\begin{center}
\setlength{\unitlength}{0.00052493in}
\begingroup\makeatletter\ifx\SetFigFont\undefined%
\gdef\SetFigFont#1#2#3#4#5{%
  \reset@font\fontsize{#1}{#2pt}%
  \fontfamily{#3}\fontseries{#4}\fontshape{#5}%
  \selectfont}%
\fi\endgroup%
{\renewcommand{\dashlinestretch}{30}
\begin{picture}(5175,4860)(0,-10)
\put(3015,4545){\blacken\ellipse{90}{90}}
\put(3015,4545){\ellipse{90}{90}}
\path(540,4545)(5040,4545)(2790,270)(540,4545)
\path(2565,3420)(2340,4545) \path(2565,3420)(3690,4545)
\path(2790,2475)(1215,4545) \path(2790,2475)(4455,4545)
\path(2790,2475)(2792,2477)(2795,2481)
    (2801,2488)(2809,2498)(2820,2512)
    (2833,2528)(2847,2546)(2862,2566)
    (2876,2588)(2890,2610)(2902,2633)
    (2913,2656)(2921,2681)(2928,2707)
    (2931,2734)(2930,2762)(2925,2790)
    (2915,2816)(2903,2839)(2889,2858)
    (2875,2872)(2862,2882)(2850,2888)
    (2838,2892)(2827,2894)(2816,2895)
    (2805,2896)(2793,2897)(2780,2900)
    (2765,2905)(2747,2913)(2727,2923)
    (2705,2937)(2680,2953)(2655,2970)
    (2628,2989)(2604,3005)(2585,3017)
    (2569,3024)(2557,3028)(2548,3029)
    (2541,3028)(2535,3026)(2529,3024)
    (2524,3024)(2517,3026)(2510,3031)
    (2501,3042)(2492,3057)(2482,3079)
    (2475,3105)(2472,3131)(2471,3156)
    (2472,3178)(2474,3197)(2477,3213)
    (2481,3226)(2485,3236)(2489,3244)
    (2494,3251)(2498,3259)(2504,3268)
    (2509,3278)(2516,3292)(2523,3310)
    (2532,3331)(2542,3358)(2553,3388)
    (2565,3420)(2579,3456)(2593,3487)
    (2605,3512)(2616,3531)(2625,3544)
    (2634,3553)(2641,3558)(2648,3562)
    (2654,3566)(2661,3572)(2667,3580)
    (2675,3592)(2682,3609)(2689,3632)
    (2696,3659)(2700,3690)(2701,3724)
    (2699,3753)(2695,3774)(2690,3789)
    (2685,3798)(2680,3803)(2674,3806)
    (2668,3809)(2661,3814)(2653,3823)
    (2645,3837)(2634,3857)(2623,3884)
    (2610,3915)(2597,3946)(2586,3974)
    (2575,3996)(2567,4013)(2559,4026)
    (2552,4035)(2546,4043)(2540,4050)
    (2535,4058)(2530,4069)(2525,4082)
    (2521,4100)(2519,4120)(2520,4140)
    (2524,4155)(2530,4167)(2536,4176)
    (2542,4181)(2547,4184)(2551,4186)
    (2554,4186)(2558,4185)(2561,4184)
    (2566,4183)(2573,4182)(2582,4182)
    (2594,4183)(2611,4183)(2631,4184)
    (2655,4185)(2681,4184)(2706,4182)
    (2728,4179)(2746,4175)(2761,4171)
    (2774,4166)(2784,4160)(2794,4155)
    (2803,4150)(2813,4145)(2825,4140)
    (2840,4137)(2858,4135)(2879,4134)
    (2902,4136)(2925,4140)(2946,4147)
    (2963,4154)(2975,4162)(2984,4168)
    (2989,4174)(2992,4178)(2993,4182)
    (2993,4185)(2992,4189)(2992,4193)
    (2993,4200)(2995,4208)(2998,4220)
    (3003,4235)(3009,4254)(3015,4275)
    (3020,4301)(3024,4327)(3026,4351)
    (3027,4374)(3026,4395)(3025,4416)
    (3023,4435)(3021,4453)(3019,4469)
    (3018,4482)(3016,4492)(3015,4497)(3015,4500)
\put(0,4545){\makebox(0,0)[lb]{\smash{{{\SetFigFont{8}{9.6}{\rmdefault}{\mddefault}{\updefault}$[0,0]$}}}}}
\put(5175,4545){\makebox(0,0)[lb]{\smash{{{\SetFigFont{8}{9.6}{\rmdefault}{\mddefault}{\updefault}$[1,1]$}}}}}
\put(2700,0){\makebox(0,0)[lb]{\smash{{{\SetFigFont{8}{9.6}{\rmdefault}{\mddefault}{\updefault}$[0,1]$}}}}}
\put(990,4680){\makebox(0,0)[lb]{\smash{{{\SetFigFont{8}{9.6}{\rmdefault}{\mddefault}{\updefault}$[r,r]$}}}}}
\put(2745,4635){\makebox(0,0)[lb]{\smash{{{\SetFigFont{8}{9.6}{\rmdefault}{\mddefault}{\updefault}$[x,x]$}}}}}
\put(2070,4635){\makebox(0,0)[lb]{\smash{{{\SetFigFont{8}{9.6}{\rmdefault}{\mddefault}{\updefault}$[r',r']$}}}}}
\put(3465,4635){\makebox(0,0)[lb]{\smash{{{\SetFigFont{8}{9.6}{\rmdefault}{\mddefault}{\updefault}$[s',s']$}}}}}
\put(4320,4635){\makebox(0,0)[lb]{\smash{{{\SetFigFont{8}{9.6}{\rmdefault}{\mddefault}{\updefault}$[s,s]$}}}}}
\put(2565,3195){\makebox(0,0)[lb]{\smash{{{\SetFigFont{8}{9.6}{\rmdefault}{\mddefault}{\updefault}$[r',s']$}}}}}
\put(2700,2295){\makebox(0,0)[lb]{\smash{{{\SetFigFont{8}{9.6}{\rmdefault}{\mddefault}{\updefault}$[r,s]$}}}}}
\put(3240,630){\makebox(0,0)[lb]{\smash{{{\SetFigFont{9}{10.8}{\rmdefault}{\mddefault}{\updefault}$\idom
= \{ [r,s]\mid 0\leq r\leq s\leq 1\}$}}}}}
\put(3600,1215){\makebox(0,0)[lb]{\smash{{{\SetFigFont{9}{10.8}{\rmdefault}{\mddefault}{\updefault}interval
domain \cite{scott72}}}}}}
\end{picture}
}
\end{center} \caption{A
schematic description of the interval domain and its order:
$[r,s]\leq [r',s']$ iff ($r\leq r'$ and $s'\leq
s$).\label{fig:idom}}
\end{figure}
The interval domain nicely illustrates some of the properties we
expect our domain model $\univ$ to have.

\begin{enumerate}
\item {\bf Refinement is complete for implementations:} Real
numbers $x\in [0,1]$ represented as intervals $[x,x]$ are the
``implementations'' of intervals, so $[r,s]$ has all $[x,x]$ with
$x\in [r,s]$ as implementations. One can easily see that $[r,s]$
is refined by $[r',s']$ iff all implementations of $[r',s']$ are
also implementations of $[r,s]$.

\item {\bf Universality:} The interval domain $\idom$ is universal
for worst/best-case abstractions of subsets of $[0,1]$. If we
abstract $X\s [0,1]$ by the interval $[\bigwedge X,\bigvee
X]\in\idom$, any element of $\idom$ is the abstraction of at least
one such $X$. In fact, there is a Galois connection
$\alpha\colon\ps {[0,1]}^{\rm op}\to \idom$ and $\gamma\colon
\idom\to\ps {[0,1]}^{\rm op}$ where $\alpha (X) = [\bigwedge
X,\bigvee X]$ is the monotone abstraction function, $\gamma
([r,s]) = [r,s]$ is the monotone ``concretization'' function, and
$\alpha\circ\gamma = {\rm id}_\idom$ and $\gamma\circ\alpha \leq
{\rm id}_{\ps {[0,1]}^{\rm op}}$.

\item {\bf Full abstraction:} The order on $\idom$ coincides with
the refinement relation as the latter means reverse containment of
implementations by item~(1) above.

\item {\bf Classical space as maximal-points space:} The set
$[0,1]$ equipped with the compact Euclidean topology is isomorphic
as a topological space to the set of maximal elements of $\idom$
in the topology induced by the Scott- or Lawson-topology of
$\idom$.

\item {\bf Denseness of computable structures:} Intervals with
rational endpoints approximate intervals to any degree of
precision.

\item {\bf Consistency measure:} The map $c\colon \idom\times
\idom \to \idom \cup \{\bot\}$ defined by $c([r,s],[r',s']) =
[\max(r,r'),\min(s,s')]$, where $[x,y]$ is understood to be $\bot$
if $x\not\leq y$, tells us whether its inputs are consistent with
each other by checking whether its output is different from
$\bot$. Non-overlapping intervals cannot possibly approximate the
same real number.
\end{enumerate}

\noindent The domain model $\univ$ for refinement of modal
transition systems \cite{hjs04} has similar properties which we
discuss briefly here prior to their technical development in this
paper. The completeness proof for implementations for refinement
of modal transition systems does not depend on the compactness of
$\max(\univ)$, is non-trivial, and presented elsewhere
\cite{huth_refinement04}. Universality amounts to showing that
every modal transition system has a refinement-equivalent
embedding in the domain $\univ$. Full abstraction means that the
order on $\univ$ equals the greatest refinement relation on
$\univ$ interpreted as a modal transition system. The
maximal-points space $\max(\univ)$ of $\univ$ gives us a precise
model of labelled transition systems and their notion of
``nearness.'' This space turns out to be the quotient of labelled
transition systems with respect to bisimulation such that the
familiar metric based on tests expressed in Hennessy-Milner logic
\cite{nicola84} induces the topology on that space. Finite-state
labelled transition systems are shown to be dense in this space.
Finally, the compactness of this space is proved and a monotone
consistency measure

\begin{equation}
c\colon \univ\times\univ\to \idom
\end{equation}

\noindent between two modal transition systems is then derived
thereof. Said compactness then renders a Galois connection between
compact sets of implementations and Scott-closed sets of modal
transition systems as shown in Theorem~\ref{theorem:gc} below.
Apart from these similarities with $\idom$, a key difference is
that $\univ$ is algebraic and that the maximal-points space is
therefore zero-dimensional.

\subsection{The domain model for refinement of modal transition
systems} The reader familiar with domain theory \cite{abramsky94}
may safely skip the next definition.

\begin{defi}\hfill 
\begin{enumerate}
\item \begin{itemize} \item A \emph{topological space} $(X,\tau)$
consists of a set $X$ and a family $\tau$ of subsets of $X$ such
that $\{\}$ and $X$ are in $\tau$, and $\tau$ is closed under
finite intersections and arbitrary unions. \item Elements $O\in
\tau$ are \emph{$\tau$-open}, complements $X\setminus O$ with
$O\in \tau$ are \emph{$\tau$-closed}, and sets that are
$\tau$-open and $\tau$-closed are \emph{$\tau$-clopen}.
\end{itemize}

\item \begin{itemize} \item A subset $A$ of a partial order
$(D,\leq)$ is \emph{directed} iff (for all $a,a'\in A$ there is
some $a''\in A$ with $a,a'\leq a''$). \item A partial order
$(D,\leq)$ is a \emph{dcpo} iff all its directed subsets $A$ have
a least upper bound $\bigvee A$. \item We write
\[\ub A\ = \{ u \in D\mid \forall a\in A\colon a\leq u\}\]
\noindent for the set of \emph{upper bounds} of $A$. \item We
denote by \[\mub A\ = \{u\in \ub A\mid \forall u'\in\ub A\colon
u'\leq u \Rightarrow u = u'\}\] \noindent the set of \emph{minimal
upper bounds} of $A$. \item An element $k\in D$ is \emph{compact}
in a dcpo $D$ iff (for all directed sets $A$ of $D$ with $k\leq
\bigvee A$ there is some $a\in A$ with $k\leq a$). We write $\ce
D$ for the set of compact elements of $D$. \item A dcpo $D$ is
\emph{algebraic} iff for all $d\in D$ the set $\{k\in\ce D\mid
k\leq d\}$ is directed with least upper bound $d$. \item For a
finite subset $F$ of $D$ define, for all $n\geq 1$

\begin{eqnarray*}
\mubi 1F &=& \mub F\\
\mubi {n+1}F &=& \mub {\mubi nF}\\
\mubi \infty F &=& \bigcup _{n\geq 1}\mubi nF\, . \end{eqnarray*}

\item A \emph{bifinite domain}, also known as an {\it SFP}-domain,
is an algebraic dcpo $D$ such that for every finite subset $F\s
\ce D$ the set $\mubi \infty F$ is finite, contained in $\ce D$,
and $\ub F = \upp {\mub F}{}$ where for any $X\s D$ we write

\[\ \
\upp X{} = \{ d\in D\mid \exists x\in X\colon x\leq d\}\\
\qquad\qquad\down X{} = \{ d\in D\mid \exists x\in X\colon d\leq
x\}
\]

\item We call $X$ \emph{upper} iff $X = \upp X{}$; \emph{lower}
iff $X = \down X{}$.\end{itemize}

\item For a bifinite domain $D$, we define \begin{itemize} \item
the \emph{Scott-topology} $\scott D$ to consist of all subsets $U$
of $D$ satisfying \[U = \upp {(U\cap \ce D)}{}\] \item the
\emph{Lawson-topology} $\lawson D$ to consist of all subsets $V$
of $D$ such that $x\in V$ implies the existence of some $k,l\in
\ce D$ with $x\in \upp k{}\setminus \upp l{}\s V$; and \item the
\emph{$\scott D$-compact saturated} subsets of $D$ to be the
$\lawson D$-closed upper subsets of $D$.
\end{itemize}
\end{enumerate}
\end{defi}

The definitions of item~(3) above are really
characterizations~\cite{abramsky94}. We use the \emph{initial}
solution $\univ$ of a domain equation, presented in \cite{hjs04}
and denoted by $\mts D$ in loc.\ cit., as the domain whose set of
maximal points we prove to be the Stone space of pointed labelled
transition systems modulo bisimulation. The items~(2) and~(3) of
Definition~\ref{def:mtsd} below are Definition~8 and~9 of
\cite{hjs04}, respectively.

\begin{defi}[\cite{hjs04}]\label{def:mtsd}\hfill 
\begin{enumerate}

\item The \emph{mixed powerdomain} \(\mpd D\)
\cite{heckmann90,gunter92a} of a bifinite domain $D$ has as
elements all pairs $(L,U)$ where $L$ is $\scott D$-closed and $U$
is $\scott D$-compact saturated such that $L$ and $U$ satisfy the
mix condition

\begin{equation}\label{equ:mix}
L = \down {(L\cap U)}{}\ .
\end{equation}

\noindent The order on $\mpd D$ is defined by

\begin{equation}
(L,U)\leq (L',U')\qquad\mbox{iff}\qquad (L\s L'\hbox{ and }U'\s
U)\, .
\end{equation}

\item Since $\mpd D$ is a bifinite domain whenever $D$ is bifinite
and since the functors ${\mathcal M}$ and $\prod$ are locally
continuous \cite{heckmann90,abramsky94}, we can solve the domain
equation

\begin{equation}\label{equ:recursion}
D = \prod _{\alpha\in\act}\mpd D
\end{equation}

\noindent over bifinite domains where $\prod _{\alpha\in\act}$
denotes the product functor over all events in $\act$, and write
$\univ$ for the \emph{initial solution} of that equation.

\item The domain $\univ$ may be \emph{interpreted as a pointed
mixed transition system}

\begin{equation}
\mts D = (\univ,\modeup \Real a,\modeup \Real c)
\end{equation}

\noindent where the recursion \(d = ((\lowpart d\alpha,\uppart
d\alpha))_{\alpha\in\act}\) of the equation~(\ref{equ:recursion})
for $\univ$ specifies that all elements $d'$ in the set $\lowpart
d\alpha$ ($\uppart d\alpha$) are exactly the $\modeup \Real
a$-successors ($\modeup \Real c$-successors) of $d$ for $\alpha$
in $\mts D$ (respectively).
\end{enumerate}
\end{defi}

Thus, the $L$ and $U$ in~(\ref{equ:mix}) model $\modeup \Real a$-
and $\modeup \Real c$-transitions within $\mts D$, respectively.
The order-theoretic mix condition~(\ref{equ:mix}) has an
equivalent version for mixed transition systems.

\begin{defi}[\cite{hjs04}]
A mixed transition system $M = (\Sigma,\modeup Ra,\modeup Rc)$
satisfies the \emph{mix condition (MC)} iff (for all
$(s,\alpha,s')\in \modeup Ra$ there is some $(s,\alpha,s'')\in
\modeup Ra\cap \modeup Rc$ such that $s'\refine{} s''$).
\end{defi}

 As shown in Proposition~3 in
\cite{hjs04},~(\ref{equ:mix}) ensures that $\mts D$ satisfies the
mix condition~(MC) since the order on $\univ$ is a refinement
within $\univ$: for all $(e,\alpha,e')\in\modeup \Real a$ there is
some $(e,\alpha,e'')\in\modeup \Real a\cap \modeup \Real c$ such
that $(\mts D, e')\refine {} (\mts D, e'')$.

\begin{exa}\label{example:mix}
Figure~\ref{fig:mixed} demonstrates that mixed transition systems
$(\Sigma,\modeup Ra,\modeup Rc$) that satisfy the mix condition
(MC) are refinement-equivalent to
 modal transition systems $(\Sigma,\modeup Ra\cap
\modeup Rc,\modeup Rc)$. Therefore, such mixed transition systems
are merely modal transition systems in disguise \cite{hjs04}.
\end{exa}
\begin{figure}
\begin{center}
\setlength{\unitlength}{0.00052493in}
\begingroup\makeatletter\ifx\SetFigFont\undefined%
\gdef\SetFigFont#1#2#3#4#5{%
  \reset@font\fontsize{#1}{#2pt}%
  \fontfamily{#3}\fontseries{#4}\fontshape{#5}%
  \selectfont}%
\fi\endgroup%
{\renewcommand{\dashlinestretch}{30}
\begin{picture}(5488,2457)(0,-10)
\put(1362,180){\ellipse{136}{136}}
\put(462,1530){\ellipse{136}{136}}
\put(2262,1530){\ellipse{136}{136}}
\put(4512,180){\ellipse{136}{136}}
\put(5412,1508){\ellipse{136}{136}}
\put(3612,1530){\ellipse{136}{136}} \path(1317,225)(507,1440)
\blacken\path(598.526,1356.795)(507.000,1440.000)(548.603,1323.513)(553.595,1370.108)(598.526,1356.795)
\path(1473,287)(2238,1367)
\blacken\path(2193.119,1251.737)(2238.000,1367.000)(2144.157,1286.418)(2189.447,1298.454)(2193.119,1251.737)
\dashline{60.000}(1362,360)(2127,1440)
\blacken\path(2082.119,1324.737)(2127.000,1440.000)(2033.157,1359.418)(2078.447,1371.454)(2082.119,1324.737)
\dashline{60.000}(417,1575)(12,1980)(462,2430)(462,1755)
\blacken\path(432.000,1875.000)(462.000,1755.000)(492.000,1875.000)(462.000,1839.000)(432.000,1875.000)
\dashline{60.000}(2172,1530)(597,1530)
\blacken\path(717.000,1560.000)(597.000,1530.000)(717.000,1500.000)(681.000,1530.000)(717.000,1560.000)
\path(4546,278)(5311,1358)
\blacken\path(5266.119,1242.737)(5311.000,1358.000)(5217.157,1277.418)(5262.447,1289.454)(5266.119,1242.737)
\dashline{60.000}(5277,1530)(3702,1530)
\blacken\path(3822.000,1560.000)(3702.000,1530.000)(3822.000,1500.000)(3786.000,1530.000)(3822.000,1560.000)
\dashline{60.000}(3522,1530)(3117,1935)(3567,2385)(3567,1710)
\blacken\path(3537.000,1830.000)(3567.000,1710.000)(3597.000,1830.000)(3567.000,1794.000)(3537.000,1830.000)
\put(192,1305){\makebox(0,0)[lb]{\smash{{{\SetFigFont{8}{9.6}{\rmdefault}{\mddefault}{\updefault}$s'$}}}}}
\put(2307,1305){\makebox(0,0)[lb]{\smash{{{\SetFigFont{8}{9.6}{\rmdefault}{\mddefault}{\updefault}$s''$}}}}}
\put(1182,1620){\makebox(0,0)[lb]{\smash{{{\SetFigFont{8}{9.6}{\rmdefault}{\mddefault}{\updefault}$\beta$}}}}}
\put(552,2025){\makebox(0,0)[lb]{\smash{{{\SetFigFont{8}{9.6}{\rmdefault}{\mddefault}{\updefault}$\beta$}}}}}
\put(1857,585){\makebox(0,0)[lb]{\smash{{{\SetFigFont{8}{9.6}{\rmdefault}{\mddefault}{\updefault}$\alpha$}}}}}
\put(1092,0){\makebox(0,0)[lb]{\smash{{{\SetFigFont{8}{9.6}{\rmdefault}{\mddefault}{\updefault}$s$}}}}}
\put(687,630){\makebox(0,0)[lb]{\smash{{{\SetFigFont{8}{9.6}{\rmdefault}{\mddefault}{\updefault}$\alpha$}}}}}
\put(1452,900){\makebox(0,0)[lb]{\smash{{{\SetFigFont{8}{9.6}{\rmdefault}{\mddefault}{\updefault}$\alpha$}}}}}
\put(4152,1575){\makebox(0,0)[lb]{\smash{{{\SetFigFont{8}{9.6}{\rmdefault}{\mddefault}{\updefault}$\beta$}}}}}
\put(4872,630){\makebox(0,0)[lb]{\smash{{{\SetFigFont{8}{9.6}{\rmdefault}{\mddefault}{\updefault}$\alpha$}}}}}
\put(3612,1980){\makebox(0,0)[lb]{\smash{{{\SetFigFont{8}{9.6}{\rmdefault}{\mddefault}{\updefault}$\beta$}}}}}
\put(3297,1350){\makebox(0,0)[lb]{\smash{{{\SetFigFont{8}{9.6}{\rmdefault}{\mddefault}{\updefault}$s'$}}}}}
\put(4242,45){\makebox(0,0)[lb]{\smash{{{\SetFigFont{8}{9.6}{\rmdefault}{\mddefault}{\updefault}$s$}}}}}
\put(5367,1260){\makebox(0,0)[lb]{\smash{{{\SetFigFont{8}{9.6}{\rmdefault}{\mddefault}{\updefault}$s''$}}}}}
\end{picture}
}
\end{center}
\caption{On the left: a mixed transition system $(\Sigma,\modeup
Ra,\modeup Rc)$ satisfying the mix condition (MC). Dashed lines
denote elements of $\modeup Rc$ and solid lines denote elements of
$\modeup Ra$. For $(s,\alpha,s')\in\modeup Ra$ there is
$(s,\alpha,s'')\in\modeup Ra\cap\modeup Rc$ with $s'\refine {}
s''$. The other tuple in $\modeup Ra$ is matched by itself as it
is in $\modeup Ra\cap \modeup Rc$. On the right: a modal
transition system that is refinement-equivalent to the mixed
transition system on the left. Its set of must-transitions is
$\modeup Ra\cap \modeup Rc$ (solid lines) and its set of
may-transitions is $\modeup Rc$ (solid or dashed lines).
 \label{fig:mixed}}
\end{figure}

\begin{rem}
By Proposition~1 in \cite{hjs04} and as seen in the previous example,
the mix condition~(MC) guarantees that the \emph{mixed} transition
system $(\univ,\modeup \Real a,\modeup \Real c)$ is
refinement-equivalent to the \emph{modal} transition system
$(\univ,\modeup \Real a\cap\modeup \Real c,\modeup \Real c)$.
Therefore all reasoning that is invariant under refinement
equivalence, as is the case in this paper, may be done with the
latter modal transition system and we abuse notation to refer to
that modal transition system as $\mts D$ as well.
\end{rem}

The domain model $\univ$ is \emph{universal}: There is an
embedding $(M,i)\mapsto \embed {M,i}$ from the class of
image-\-finite pointed mixed transition system satisfying the
mix-condition~(MC) to elements of $\univ$ such that $(M,i)$ and
$(\mts D,\embed {M,i})$ are refinement-equivalent (Theorem~6.1 in
\cite{hjs04}). The domain model $\univ$ is \emph{fully abstract}:
For all $d,e\in\univ$, we have $d\leq e$ iff $(\mts D,d)\refine {}
(\mts D,e)$ (Theorem~5 in \cite{hjs04}). For sake of completeness,
we sketch the construction of this embedding and needed aspects of
the full abstraction proof in the next section.

\section{Stone space of labelled transition
systems}\label{section:stone} We show that the maximal elements of
$\univ$ are precisely the representations of pointed labelled
transition systems modulo bisimulation; and that this quotient is
a Stone space and therefore determined by a complete ultra metric.

\subsection{The maximal-points space}

We define the required notions from topology.

\begin{defi}\label{def:ultra}\hfill 
\begin{enumerate}
\item A topological space $(X,\tau)$ is

\begin{enumerate}
\item \emph{compact} iff for all ${\mathcal U}\s\tau$ with $X\s
\bigcup {\mathcal U}$ there is a finite subset ${\mathcal F}\s
{\mathcal U}$ with $X\s \bigcup {\mathcal F}$;

\item \emph{Hausdorff} iff for all $x\not= x'$ in $X$ there are
$O,O'\in \tau$ with $x\in O$, $x'\in O'$ and $O\cap O' =\{\}$;

\item \emph{zero-dimensional} iff every $\tau$-open set is the
union of $\tau$-clopens; and

\item a \emph{Stone space} iff it is zero-dimensional, compact,
and Hausdorff.
\end{enumerate}

\item A subset $C$ of $(X,\tau)$ is \emph{$\tau$-compact} iff the
topological space $(C, \{U\cap C\mid U\in\tau\})$ is compact.

\item A subset $A$ of $X$ is \emph{dense} in $(X,\tau)$ iff $A\cap
O$ is non-empty for all non-empty $O\in\tau$.

\item\label{item:fourd} An \emph{ultra-metric} on $X$ is a
function $d\colon X\times X\to [0,1]$ such that for all $x,y,z\in
X$

\begin{enumerate}
\item $d(x,y) = 0$ iff $x = y$; \item $d(x,y) = d(y,x)$; and \item
$d(x,z)\leq \max (d(x,y),d(y,z))$.
\end{enumerate}

\item An ultra-metric $d\colon X\times X\to [0,1]$
\emph{determines a topology $\tau _d$} on $X$ whose elements are
all those $O\s X$ that are unions of sets of the form $\ball \eta
x = \{y\in X\mid d(x,y) < \eta\}$ for $x\in X$ and rational $\eta
> 0$.

\item A topological space $(X,\tau)$ is \emph{ultra-metrizable}
iff there is an ultra-metric $d\colon X\times X\to [0,1]$ such
that $\tau = \tau _d$.

\item We denote by \(\max (\univ)\ = \{m\in \univ\mid \forall
d\in\univ\colon m\leq d \Rightarrow m = d\}\) the set of
\emph{maximal elements} of $\univ$. The set

\begin{equation}
\univx = \max (\univ)
\end{equation}

\noindent has a \emph{maximal-points space topology}
\cite{lawson97}

\begin{equation}
\tau _\univx = \{ U\cap\univx\mid U\in \scott \univ\}\, .
\end{equation}

\item For $d\in \univ$, we write

\begin{equation}
\meu d = \upp d{}\cap \max(\univ)\, .
\end{equation}

\end{enumerate}
\end{defi}

Since $\univ$ is a bifinite domain, the Lawson
condition~\cite{lawson97} holds for $\univ$, namely that the
topology $\tau _\univx$ is also induced by the
$\lawson\univ$-topology:

\begin{equation}\label{equ:lawsoncondition}
\tau _\univx = \{ V\cap\univx\mid V\in \lawson \univ\}\, .
\end{equation}

We remark that not all bifinite domains $D$ enjoy the property
that $\max (D)$ is compact in the topology induced by $\scott D$
or $\lawson D$.

\subsection{Maximal-points space is zero-dimensional and
Hausdorff} We first record that $\tau_\univx$ is Hausdorff and
zero-dimensional. Proposition~\ref{prop:zero} below holds for any
algebraic domain satisfying the Lawson condition~\cite{lawson97}.
We state and prove that proposition for $\univ$ for sake of
completeness.

\begin{prop}\label{prop:zero}
The topological space $(\univx,\tau _\univx)$ is zero-dimensional
and Hausdorff.
\end{prop}

\proof\hfill 
\begin{itemize} \item Every $U\in\scott\univ$ is the union
of $\scott \univ$-opens $\upp k{}$, $k\in\ce \univ$, as $\univ$ is
algebraic. But each $\upp k{}$ is $\lawson\univ$-clopen as
$\scott\univ\s\lawson\univ$ and $\upp k{}$ is
$\lawson\univ$-closed. From the Lawson condition for
$\univ$,~(\ref{equ:lawsoncondition}), we infer that $\meu k$ is
$\tau_\univx$-clopen and so $\tau _\univx$ is zero-dimensional as
every $O\in\tau _\univx$ is the union of such sets. \item To show
that $\tau _\univx$ is Hausdorff, let $x\not = y$. Since $\univ$
is a partial order we may assume $x\not\leq y$ without loss of
generality. Since $\univ$ is algebraic, $x\not\leq y$ implies
$k\leq y$ and $k\not\leq x$ for some $k\in\ce\univ$. But $\meu k$
is $\tau _\univx$-open and contains $y$ whereas $x$ is in
$\univx\setminus \meu k$ which is $\tau_\univx$-open since $\meu
k$ is also $\tau_\univx$-closed.
\end{itemize}
\qed

\subsection{Semantics of Hennessy-Milner logic} We use tools from temporal
logic to develop a sufficient criterion for membership in $\max
(\univ)$.

\begin{defi}\label{def:weak}\hfill 
\begin{enumerate}
\item The set of formulas of \emph{Hennessy-Milner logic}
\cite{hennessy85} is generated by the grammar

\begin{equation}
\phi ::= \true\ \mid\ \neg\phi\ \mid\ \dia\alpha\phi\ \mid\
\phi\land\phi \end{equation}

\noindent where $\alpha$ ranges over the finite set of events
$\act$.

\item Let $(N,i) = ((\Sigma, \modeup Ra, \modeup Rc),i)$ be a
pointed modal transition system. Larsen's semantics, denoted by
$\models$ in \cite{larsen89b} for Hennessy-Milner logic in
negation normal form, is depicted in Figure~\ref{fig:hml}.

\item We write $\bx\alpha$ for $\neg\dia\alpha\neg$ and
$\phi\lor\psi$ for $\neg (\neg\phi\land\neg\psi)$ subsequently for
all $\alpha\in\act$ and all $\phi$ and $\psi$ of Hennessy-Milner
logic.
\end{enumerate}
\end{defi}
\begin{figure}
\begin{center}
\[
\begin{array}{lll}
(N,i)\modeup\models m\true & {} &{}\\
(N,i)\modeup\models m\neg\phi &\mbox{iff}&
(N,i)\not\!\!\!{\modeup\models {\neg m}}\phi\\
(N,i)\modeup\models m\dia\alpha\phi & \mbox{iff} & \mbox{for
some}\
(i,\alpha,i')\in \modeup Rm,\ (N,i')\modeup\models m\phi\\
(N,i)\modeup\models m\phi\land\psi & \mbox{iff} &
((N,i)\modeup\models m\phi\ \mbox{and}\ (N,i)\modeup\models m\psi)
\end{array}
\]
\caption{Semantics of Hennessy-Milner logic with two judgments
$(N,i)\modeup\models m\phi$ where ${\it m}\in \{{\it a}, {\it
c}\}$, ${\it \neg a = c}$, and ${\it \neg c = a}$.\label{fig:hml}}
\end{center}
\end{figure}

\begin{rem}
For each ${\it m}\in \{{\it a}, {\it c}\}$ we have

\[
\begin{array}{lll}
(N,i)\modeup\models m\bx\alpha\phi &\mbox{iff} & \mbox{for all
$(i,\alpha,i')\in\modeup R{\it \neg m}$, $(N,i')\modeup\models
m\phi$}\\
(N,i)\modeup\models m\phi\lor\psi & \mbox{iff}
&\mbox{$((N,i)\modeup\models m\phi$ or $(N,i)\modeup\models
m\psi)$\, .}\end{array}
\]

\noindent Please note that $\modeup\models m\bx\alpha\phi$
universally quantifies over transitions in the \emph{dual} mode
$\neg {\it m}$.
\end{rem}

\begin{exa}\label{example:two}
Consider the modal transition system $N$ in Figure~\ref{fig:ref}.

\begin{enumerate}
\item\label{item:one} We have $(N,{\rm Talks})\modeup\models
c\dia{\rm drinks}\true$ because of the $\modeup Rc$-transition
$({\rm Talks},{\rm drinks},{\rm Drinks})$. By the semantics of
negation, this implies $(N,{\rm Talks})\not\!\!\!{\modeup\models
a}\neg\dia{\rm drinks}\true$. We also infer $(N,{\rm
Talks})\not\!\!\!{}{\modeup\models a}\dia{\rm drinks}\true$ as
there is no state $s$ with $({\rm Talks},{\rm drinks},s)\in\modeup
Ra$. By the semantics of disjunction, these two judgments render
$(N,{\rm Talks})\not\!\!\!{\modeup\models a}\dia{\rm
drinks}\true\lor\neg\dia{\rm drinks}\true$. This judgment says
that we can't determine that $\dia{\rm
drinks}\true\lor\neg\dia{\rm drinks}\true$ is asserted in state
${\rm Talks}$ in $N$. As that formula is a tautology over labelled
transition systems we see that judgments $(N,{\rm Talks})\models
^a\phi$ under-approximate validity judgments ``all refinements of
$(N,{\rm Talks})$ satisfy $\phi$.'' As we show below, it turns out
that the ability to capture these validity judgments for certain
tautologies over labelled transition systems via $\models ^a$ is
what characterizes modal transition systems that are
refinement-equivalent to labelled transition systems.

\item\label{item:two} We have $(N,{\rm
Waits})\not\!\!\!{\modeup\models a}\bx {\rm newPint}\bx{\rm
talks}(\dia{\rm drinks}\true\lor\neg\dia{\rm drinks}\true)$ as
there is an $\modeup Rc$-path $({\rm Waits},{\rm newPint},{\rm
Drinks})({\rm Drinks},{\rm talks},{\rm Talks})$ for the word ${\rm
newPint}\,{\rm talks}\in\act ^*$ and $(N,{\rm
Talks})\not\!\!\!{\modeup\models a}\dia{\rm
drinks}\true\lor\neg\dia{\rm drinks}\true$ by item~(1). Therefore,
the check $(N,{\rm Waits})\not\!\!\!{\modeup\models a}\bx {\rm
newPint}\bx{\rm talks}(\dia{\rm drinks}\true\lor\neg\dia{\rm
drinks}\true)$ is unable to validate a tautology over labelled
transition systems at state {\rm Waits} in $N$.
\end{enumerate}
\end{exa}

\subsection{Denseness of image-finite labelled transition systems}
We sketch the definition of the embedding $\embed {M,i}\in \univ$
for an image-finite modal transition system $(M,i)$ such that
$(M,i)$ and $(\mts D,\embed {M,i})$ are refinement-equivalent
\cite{hjs04}. This construction follows ideas from algebraic
semantics \`a la Nivat-Courcelle-Guessarian \cite{CN76} or \`a la
Goguen-Thatcher-Wagner-Wright \cite{GWWT77} in that we unfold
pointed modal transition systems as finite trees for a fixed
depth, adding a may-stub to all leaves of that tree for which
there are still outgoing transitions in the pointed modal
transition system. This unfolding is presented here via a simple
process algebra.

\begin{defi}\hfill 
\begin{enumerate}
\item The grammar for the \emph{process algebra $\mpa$} is

\begin{equation}
p ::= \ {\bf 0}\ \mid\ {\bf \bot}\ \mid\ \alpha_\true.p\ \mid\
\alpha_\bot.p\ \mid\ p + p
\end{equation}

\noindent where $\alpha$ ranges over the finite set of events
$\act$ and no $p$ in $p + p$ is allowed to be ${\bf \bot}$ or
${\bf 0}$.

\item For each $p\in\mpa$ let $\denoteD p\in\univ$ be as in
Figure~\ref{fig:paD}.

\item For all $p\in\mpa$, the structural operational semantics in
Figure~\ref{fig:sos} defines a pointed modal transition system
$(\denote p{}, p)$.
\end{enumerate}
\end{defi}
\begin{figure}
\begin{center}
\begin{tabular}{l}
$\denoteD {{\bf 0}} = ((\{\},\{\}))_{\alpha\in\act}$ \\\\[0.2ex]
$\denoteD
\bot = ((\{\},\univ))_{\alpha\in\act}$ \\\\[0.2ex]
$(\lowpart {\denoteD {\alpha_\true.p}}\alpha,\uppart {\denoteD
{\alpha_\true.p}}\alpha) = (\down {\denoteD p}{},\upp {\denoteD
p}{}{})$ \\\\[0.2ex] $(\lowpart {\denoteD {\alpha_\true.p}}\beta,\uppart
{\denoteD {\alpha_\true.p}}\beta) = (\{\},\{\}),\
\alpha\not=\beta$
 \\\\[0.2ex]
 $(\lowpart {\denoteD {\alpha_\bot.p}}\alpha,\uppart {\denoteD
{\alpha_\bot.p}}\alpha) = (\{\},\upp {\denoteD p}{}{})$ \\\\[0.2ex]
$(\lowpart {\denoteD {\alpha_\bot.p}}\beta,\uppart {\denoteD
{\alpha_\bot.p}}\beta) = (\{\},\{\}),\ \alpha\not=\beta$
 \\\\[0.2ex]
 ${\denoteD {p + q}}^{\it m}_\gamma = {\denoteD p}^{\it m}_\gamma
 \cup {\denoteD q}^{\it m}_\gamma,\ \gamma\in\act,\ m\in \{ a,c\}$ \\\\
\end{tabular}
\end{center}
\caption{A denotational semantics of $\mpa$ in $\univ$ that
interprets ${\bf 0}$ as deadlock, $\bot$ as the least element, $+$
as the mix union of \cite{heckmann90}, and the prefixes as
expected using saturations with $\downarrow$ and $\uparrow$ to
ensure membership in $\univ$. \label{fig:paD}}
\end{figure}
\begin{exa}
Let $p\in\mpa$ be ${\rm drinks} _\bot.\bot + {\rm orders}_\bot
.\bot + {\rm talks}_\true . {\bf 0}$. Then $(\denote p{},p)$ is
refinement-equivalent to the image-\-finite pointed modal
transition system in Figure~\ref{fig:app}.
\end{exa}

\noindent We record that the denotational semantics of $\mpa$ in
$\univ$ matches the structural operational semantics. This proof
is straightforward and amounts to showing that the saturations
with $\downarrow$ and $\uparrow$ in $\univ$ do not break
refinement equivalence as they always occur in the right
direction.

\begin{lem}[\cite{huth_refinement04}]\label{lemma:match}
For all $p\in\mpa$, the modal transition system $(\denote p{},p)$
is refinement-equivalent to the mixed transition system $(\mts
D,\denoteD p)$.
\end{lem}

To define the embedding $\embed {M,i}$ for an image-\-finite
pointed modal transition system $(M,i)$ consider $m\geq 0$, unwind
$M$ from $i$ as a tree $M[m]$ such that all, and only, paths of
length $\leq m$ of $M$ are present. If a leaf of that tree has
some $\modeup Rc$-successor in $M$, create $\modeup Rc$-loops on
that leaf for \emph{all} events in $\act$ (a \emph{may-stub});
otherwise, leave it as is (deadlock). By construction, this
image-\-finite pointed modal transition system $(M[m],i)$ is the
operational meaning $(\denote {p_m}{}, p_m)$ of a term
$p_m\in\mpa$ so $m\leq m'$ and Lemma~\ref{lemma:match} imply that
$\denoteD {p_m}\leq \denoteD {p_{m'}}$. Thus $\{ \denoteD
{p_m}\mid m\geq 0\}$ is directed and we can set

\begin{equation}
\embed {M,i} = \bigvee _{m\geq 0} \denoteD {p_m}
\end{equation}

\noindent and note, shown in \cite{heckmann90} for bifinite
domains without reference to a process algebra, that

\begin{equation}
\ce\univ = \{\denoteD p\mid p\in\mpa \}\, .
\end{equation}

\noindent We may thus represent all $k\in\ce\univ$ in the form
$\denoteD p$ for some $p\in\mpa$ subsequently.

\begin{figure}
\begin{center}

\begin{tabular}{ll}
{} & $\prt{{}}{\bot\longrightarrow^\gamma_\bot \bot}{{\rm
MayStub}}$ \\\\[0.2ex] $\prt{{}}{\alpha_\true.p\longrightarrow^\alpha_{\true}
p}{{\rm MustPrefix}}$ &
$\prt{{}}{\alpha_\bot.p\longrightarrow^\alpha_{\bot}
p}{{\rm MayPrefix}}$ \\\\[0.2ex]
$\prt{p\longrightarrow ^\alpha_v p'}{p + q\longrightarrow
^\alpha_v p'}{{\rm LChoice}}$ & $\prt{q\longrightarrow ^\alpha_v
q'}{p + q\longrightarrow ^\alpha_v q'}{{\rm RChoice}}$
\\\\
\end{tabular}

\end{center}
\caption{Structural operational semantics of $\mpa$ in $\univ$:
$p\longrightarrow^\alpha_\bot p'$ and $p\longrightarrow
^\alpha_\true p'$ denote a may-transition (respectively)
must-transition from $p$ to $p'$, with label $\alpha\in\act$.
There are no transitions out of ${\bf 0}$;  $v\in \{\bot,\true\}$;
and the occurrence of $\gamma$ ranges over all events in $\act$.
\label{fig:sos}}
\end{figure}

\begin{exa}
Figure~\ref{fig:app} illustrates the construction of a finite
approximation and depicts $(M[1],{\rm TomDrinks})$ for the pointed
modal transition system $(M,{\rm TomDrinks})$ of
Figure~\ref{fig:reftwo}.
\end{exa}

\begin{figure}
\begin{center}
\setlength{\unitlength}{0.00052493in}
\begingroup\makeatletter\ifx\SetFigFont\undefined%
\gdef\SetFigFont#1#2#3#4#5{%
  \reset@font\fontsize{#1}{#2pt}%
  \fontfamily{#3}\fontseries{#4}\fontshape{#5}%
  \selectfont}%
\fi\endgroup%
{\renewcommand{\dashlinestretch}{30}
\begin{picture}(3900,2634)(0,-10)
\dashline{60.000}(2037,225)(2037,1215)
\blacken\path(2067.000,1095.000)(2037.000,1215.000)(2007.000,1095.000)(2037.000,1131.000)(2067.000,1095.000)
\dashline{60.000}(1722,225)(912,1215)
\blacken\path(1011.207,1141.122)(912.000,1215.000)(964.770,1103.128)(965.192,1149.988)(1011.207,1141.122)
\dashline{60.000}(597,1575)(12,2070)(687,2340)(687,1755)
\blacken\path(657.000,1875.000)(687.000,1755.000)(717.000,1875.000)(687.000,1839.000)(657.000,1875.000)
\dashline{60.000}(1992,1575)(1407,2070)(2082,2340)(2082,1755)
\blacken\path(2052.000,1875.000)(2082.000,1755.000)(2112.000,1875.000)(2082.000,1839.000)(2052.000,1875.000)
\path(2307,225)(3432,1215)
\blacken\path(3361.733,1113.203)(3432.000,1215.000)(3322.096,1158.246)(3368.940,1159.507)(3361.733,1113.203)
\put(1812,0){\makebox(0,0)[lb]{\smash{{{\SetFigFont{8}{9.6}{\rmdefault}{\mddefault}{\updefault}TomDrinks}}}}}
\put(462,1350){\makebox(0,0)[lb]{\smash{{{\SetFigFont{8}{9.6}{\rmdefault}{\mddefault}{\updefault}TomDrinks}}}}}
\put(1812,1350){\makebox(0,0)[lb]{\smash{{{\SetFigFont{8}{9.6}{\rmdefault}{\mddefault}{\updefault}Waits}}}}}
\put(3162,1350){\makebox(0,0)[lb]{\smash{{{\SetFigFont{8}{9.6}{\rmdefault}{\mddefault}{\updefault}TomTalks}}}}}
\put(822,495){\makebox(0,0)[lb]{\smash{{{\SetFigFont{8}{9.6}{\rmdefault}{\mddefault}{\updefault}drinks}}}}}
\put(2307,720){\makebox(0,0)[lb]{\smash{{{\SetFigFont{8}{9.6}{\rmdefault}{\mddefault}{\updefault}talks}}}}}
\put(1407,990){\makebox(0,0)[lb]{\smash{{{\SetFigFont{8}{9.6}{\rmdefault}{\mddefault}{\updefault}orders}}}}}
\put(12,2475){\makebox(0,0)[lb]{\smash{{{\SetFigFont{8}{9.6}{\rmdefault}{\mddefault}{\updefault}$\forall\gamma\in\act$}}}}}
\put(1632,2475){\makebox(0,0)[lb]{\smash{{{\SetFigFont{8}{9.6}{\rmdefault}{\mddefault}{\updefault}$\forall\gamma\in\act$}}}}}
\end{picture}
}
\end{center}
\caption{The pointed modal transition system $(M[1],{\rm
TomDrinks})$, an approximation of the pointed modal transition
system $(M,{\rm TomDrinks})$ in Figure~\ref{fig:reftwo}. The
states {\rm Waits} and the second {\rm TomDrinks} turn into
may-stubs whereas the approximation recognizes ${\rm TomTalks}$ as
a deadlocked state.\label{fig:app}}
\end{figure}

We define the characteristic formulas for terms $p$ of the process
algebra $\mpa$, which will also be the characteristic formulas of
the compact elements $\denoteD p$ of $\univ$.

\begin{defi}
For each $p\in\mpa$, we define the formula $\phi _p$ of
Hennessy-Milner logic in Figure~\ref{fig:phip}.
\end{defi}
\begin{figure}
\begin{center}

\begin{tabular}{l}
$\phi _{\bf 0} =  \bigwedge _{\alpha\in\act}\neg\dia\alpha\true$
\\\\[3ex]
 $\phi _{\bot} = \true$ \\\\[3ex]
$\phi _{\alpha_\true.p} = \dia\alpha\phi _p\land \bx\alpha\phi
_p\land\bigwedge _{\beta\not=\alpha} \neg\dia\beta\true$ \\\\[3ex]
$\phi _{\alpha_\bot.p} = \bx\alpha\phi _p\land \bigwedge
_{\beta\not=\alpha} \neg\dia\beta\true$ \\\\[3ex] $\phi _{p+q} =
\bigwedge \{\dia\alpha\phi _{r'}\mid \alpha\in\act,\
p+q\longrightarrow ^\alpha_\true r'\}$ \\\\[3ex]
$\ \ \ \ \ \ \ \ \ \  \land \bigwedge _{\alpha\in\act}\bx\alpha
(\bigvee _{v\in \{\bot,\true \}} \{ \phi _{r'}\mid
p+q\longrightarrow ^\alpha_v r'\})$
\end{tabular}

\end{center}
\caption{The characteristic formulas $\phi_p$ for terms $p$ of the
process algebra $\mpa$.\label{fig:phip}}
\end{figure}
These formulas characterize their terms, for one can interchange
refinement checks $(\mts D, \denoteD p)\refine {} (\mts D,d)$ with
model checks $(\mts D, d)\modeup\models a\phi _p$ for all
$d\in\univ$.

\begin{lem}\label{lemma:psip}
For all $d\in\univ$ we have

\begin{equation}\label{equ:phip}
\denoteD p\leq d\qquad\mbox{iff}\qquad (\mts D, d)\modeup\models
a\phi _p\, .
\end{equation}

\end{lem}

\proof We prove this by structural induction on $p\in\mpa$.
\begin{itemize}
\item We have $\denoteD {\bf 0}\leq d$ iff (there are no
$\modeup\Real c$-transitions out of $d$) iff $(\mts
D,d)\modeup\models a\psi _{\bf 0}$.

\item We have $\denoteD \bot\leq d$ for all $d\in\univ$, but also
$(\mts D,d)\modeup\models a\psi _{\bot}$ for all $d\in\univ$.

\item Using induction on $p$, we have $(\mts D,d)\modeup\models
a\psi _{\alpha_\true.p}$ iff (there is a $\modeup \Real
a$-transition $(d,\alpha,d')$ in $\univ$ with $\denoteD p\leq d'$;
all $\modeup \Real c$-transitions $(d,\alpha,d'')$ in $\univ$
satisfy $\denoteD p\leq d''$; and there are no $\modeup \Real
c$-transitions out of $d$ in $\univ$ for other events). This
exactly captures $\denoteD {\alpha_\true.p}\leq d$.

\item By induction on $p$, we have $(\mts D,d)\modeup\models a\psi
_{\alpha_\bot.p}$ iff (there are no $\modeup \Real c$-transitions
out of $d$ for events other than $\alpha$, and all $\modeup \Real
c$-transitions $(d,\alpha,d')$ satisfy $\denoteD p\leq d'$). But
this captures $\denoteD {\alpha_\bot.p}\leq d$.

\item Let $(\mts D,d)\modeup\models a \psi _{p+q}$. Then $(\mts
D,d)\modeup\models a \bigwedge _{\alpha\in\act}\bigwedge
_{p+q\longrightarrow ^\alpha_\true r'} \dia\alpha\psi _{r'}$ and
induction express that all $\modeup \Real a$-transitions out of
$p+q$ to some $r'$ can be answered by corresponding
$(d,\alpha,d')\in \modeup\Real a$ with $\denoteD r'\leq d'$;
whereas $(\mts D,d)\modeup\models a \bigwedge
_{\alpha\in\act}\bx\alpha \bigl ( \bigvee \{ \psi _{r'}\mid\exists
v\in \{\bot,\true\}\colon p+q\longrightarrow ^\alpha_v r'\}\bigr
)$ states that all $(d,\alpha,d')\in\modeup \Real c$ can be
answered in $(\denoteD {p+q}{}, p+q)$ by corresponding $\modeup
\Real c$-transitions to some $r'$ such that $r'\leq d'$ by
induction. So $d\leq \denoteD{p+q}$.\qed
\end{itemize}

\noindent This characterization is the key to proving that $\univ$
is fully abstract and that refinement is characterized by the
semantics for Hennessy-Milner logic.

\begin{cor}[\cite{hjs04}]\label{cor:hjs04}\hfill 
\begin{enumerate}
\item The order on $\univ$ is the greatest refinement relation
within $\mts D$.

\item For all pointed modal transition systems $(M,i)$ and $(N,j)$
the following are equivalent:

\begin{enumerate}
\item $(M,i)\refine {} (N,j)$

\item for all $\phi$ of Hennessy-Milner logic,
$(M,i)\modeup\models a\phi$ implies $(N,j)\modeup\models a\phi$

\item for all $\phi$ of Hennessy-Milner logic,
$(N,j)\modeup\models c\phi$ implies $(M,i)\modeup\models c\phi$.
\end{enumerate}
\end{enumerate}
\end{cor}

\proof\hfill 
\begin{enumerate} \item That the order of $\univ$ is a
refinement follows directly from the definition of $\mts D$. For
the converse, we show ``$d\not\leq e$ implies that $(\mts D, e)$
does not refine $(\mts D,d)$:'' First note that $\ce\univ$
order-generates $\univ$ so $d\not\leq e$ implies $k\leq d$ and
$k\not\leq e$ for some $k\in\ce\univ$. Then there is $p\in\mpa$
with $k = \denoteD p$ so that, by Lemma~\ref{lemma:psip}, for all
$f\in \univ$: $k\leq f$ iff $(\mts D,f)\modeup\models a\phi _p$.
Thus, $(\mts D,d)\modeup\models a\phi _p$ and $(\mts
D,e)\not\!\!\!{\modeup\models a}\phi _p$ imply that $e$ does not
refine $d$ in $\mts D$.

\item Since $\modeup\models a$ and $\modeup\models c$ are dual
with respect to negation, (b) and (c) are equivalent. The proof
that (a) implies (b) is a straightforward structural induction on
$\phi$ \cite{hjs01}. That (b) implies (a) follows from item~(1),
Lemma~\ref{lemma:psip}, and the fact that $\univ$ is
algebraic.\qed
\end{enumerate}

We demonstrate that embeddings of pointed image-\-finite labelled
transition systems are dense in $(\univx, \tau _{\univx})$, which
we subsequently show to be the quotient space of all pointed
labelled transition systems with respect to bisimulation. The
denseness argument rests on the fact that embeddings of
implementations are maximal elements of $\univ$.

\begin{lem}\label{lemma:aremax}
Let $d\in\univ$ be such that, for all $\phi$ of Hennessy-Milner
logic, $(\mts D,d)\modeup\models c\phi$ implies $(\mts
D,d)\modeup\models a\phi$. Then $d\in\max (\univ)$.
\end{lem}

\proof Consider such a $d$ and let $d\leq e$ in $\univ$. Since
$\leq$ is a partial order and since $\univ$ is algebraic it
suffices to show that $\down e{}\cap \ce \univ\s \down d{}$. So
let $\denoteD p\in \ce \univ$ with $\denoteD p\leq e$. For $\phi
_p$ of~(\ref{equ:phip}), $\denoteD p\leq e$ implies $(\mts
D,e)\modeup\models a\phi _p$ which implies $(\mts
D,e)\modeup\models c\phi _p$ by Corollary~\ref{cor:hjs04} as
$\univ$ is fully abstract. But $d\leq e$ means $(\mts D, d)\refine
{}(\mts D, e)$ as $\univ$ is fully abstract, and so $(\mts
D,d)\modeup\models c\phi _p$ by Corollary~\ref{cor:hjs04} as
$(\mts D,e)\modeup\models c\phi _p$. By assumption on $d$, this
renders $(\mts D,d)\modeup\models a\phi _p$ and so $\denoteD p\leq
d$ by~(\ref{equ:phip}).\qed

\begin{prop}\label{prop:dense}
The set of all embeddings of pointed image-\-finite labelled
transition systems is dense in $(\univx, \tau _\univx)$.
\end{prop}

\proof As any pointed image-\-finite labelled transition system
$(L,l)$ is refinement-equivalent to $(\mts D,\embed{L,l})$
\cite{hjs04}, the embedding $\embed {L,l}$ is in
$\max(\univ)=\univx$ since it satisfies the assumptions of
Lemma~\ref{lemma:aremax}.

Let $O\in\tau _\univx$ be non-empty, so $O = U\cap \max (\univ)$
for some $U\in\scott\univ$ and there is some $k\in \ce\univ$ with
$\meu k\s U\cap \max (\univ)$ since $O$ is non-empty and $\univ$
is algebraic. Let $q\in\mpa$ be obtained by replacing all $\bot$
in $p$ with ${\bf 0}$ and, for all $\gamma\in\act$, all prefixes
$\gamma _\bot.$ with $\gamma _\true$. Then $(\denote q{}, q)$
refines $(\denote p{}, p)$. Since $(\denote q{},q)$ is a pointed
\emph{labelled} transition system and $(\denote r{}, r)$ is
refinement-equivalent to $(\mts D, \denoteD r)$ for all $r\in\mpa$
by Lemma~\ref{lemma:match}, we conclude $\denoteD q \in\meu k\s O$
by Lemma~\ref{lemma:aremax} and $\denoteD q$ is the embedding of a
pointed image-\-finite labelled transition system. \qed

\subsection{Compactness of maximal-points space}
We show that $(\univx,\tau _\univx)$ is compact by proving,
indirectly, that $\max (\univ)$ is $\lawson\univ$-closed. Using
results from \cite{alessi03} one could show that $\max (\univ)$ is
$\lawson\univ$-closed by finding a subset $T$ of $\ce \univ$ that
is a finitely branching tree and co-final in $\ce\univ$. Given a
candidate of such a $T$, the property that is difficult to
ascertain is that any two elements of $T$ that have an upper bound
in $\ce\univ$ are comparable. For example, consider the compact
elements $\denoteD {\alpha_\true.\bot + \beta_\true.{\bf 0}}$ and
$\denoteD {\alpha_\true.{\bf 0} + \beta_\true.\bot}$, both of
which have the compact element $\denoteD {\alpha_\true.{\bf 0} +
\beta_\true.{\bf 0}}$ as an upper bound yet neither of them
refines the other.

Faced with these difficulties, we therefore take a different route
and realize $\max (\univ)$ as the set of those elements $d$ of
$\univ$ that pass a set of judgments $(\mts D, d)\modeup\models
a\formula w\alpha p$ where $\formula w\alpha p$ are formulas of
Hennessy-Milner logic.

\begin{defi}\hfill 
\begin{enumerate}
\item Let $w = \delta _1\delta _2\dots\delta _n\in\act ^*$,
$\alpha\in\act$, and $p\in\mpa$. Then we define the
Hennessy-Milner logic formula

\begin{equation}
\formula {w}\alpha p = \bx {\delta _1}\bx {\delta _2}\dots \bx
{\delta _n}(\dia \alpha \phi _p\lor \neg \dia \alpha \phi _p)
\end{equation}

\noindent with $\phi _p$ as in Figure~\ref{fig:phip}.

\item Let $\Phi$ be the set of all Hennessy-Milner logic formulas
$\formula w\alpha p$ where $w\in\act ^*$, $\alpha\in\act$, and
$p\in\mpa$.

\item For $\phi$ of Hennessy-Milner logic and all ${\it m}\in
\{{\it a}, {\it c}\}$ we define

\begin{equation}\denotemode \phi m
= \{ d\in \univ\mid (\mts D,d)\modeup\models m\phi \}\, .
\end{equation}

\item Let $C_\Phi = \bigcap _{\phi\in\Phi} \denotemode \phi a$.
\end{enumerate}
\end{defi}

For each formula $\phi$ in $\Phi$, the test $(\mts D,
d)\modeup\models a\phi$ checks whether there is a certain $\modeup
\Real c$-reachable state from $d$ with a certain outgoing
may-transition that cannot be matched with a corresponding
outgoing must-transition. Accordingly, $C_\Phi$ consists of those
elements whose reachable states always find such a match.
Intuitively, those should be the elements that represent labelled
transition systems.

\begin{exa} The formulas in items~(\ref{item:one})
and~(\ref{item:two}) of Example~\ref{example:two} are in $\Phi$ as
$\true$ is $\phi _{\bot _\univ}$, $\denoteD \bot = \bot
_{\univ}\in\ce\univ$, and $\epsilon\in\act ^*$.
\end{exa}

Rather than proving directly that $\max (\univ)$ is
$\lawson\univ$-closed, we first establish that $C_\Phi$ is
$\lawson\univ$-closed and then prove $\max (\univ) = C_\Phi$.
Whence maximal elements in $\univ$ are exactly those elements
whose reachable may-transitions have matching must-transitions. As
$C_\Phi$ is the intersection of sets of the form $\denotemode \phi
a$, we can show that the former is $\lawson\univ$-closed by
proving that all latter sets are $\lawson\univ$-closed. We do this
by structural induction on $\phi$ which requires a stronger
induction hypothesis.

\begin{lem}\label{lemma:denoteclosed}
For each $\phi$ of Hennessy-Milner logic, the sets $\denotemode
\phi a$ and $\denotemode \phi c$ are $\lawson\univ$-clopen. In
particular, $C_\Phi$ is $\lawson\univ$-closed.
\end{lem}

\proof We proceed with the first claim by structural induction on
$\phi$. This is evident for the clauses $\true$, negation, and
conjunction since $\denotemode \true m = \univ$ is
$\lawson\univ$-clopen and clopens are closed under set complement
($\denotemode {\neg\phi} a = \univ \setminus \denotemode \phi c$
and $\denotemode {\neg\phi}c = \univ \setminus \denotemode \phi
a$) and finite intersections. We still require proofs for
$\dia\alpha\phi$, where for each ${\it m}\in \{{\it a}, {\it c}\}$
we have

\begin{equation}\label{equ:diauniv}
\denotemode{\dia\alpha\phi}m = \{d\in \univ\mid \modeup dm_\alpha
\cap \denotemode\phi m\not=\e\}\, .
\end{equation}

  \begin{itemize}
  \item Let ${\it m} = {\it a}$. By Theorem~4.2 in~\cite{hjs04}, $\denotemode\psi a$
   is $\scott\univ$-open for all $\psi$ of Hennessy-Milner logic, so
  $\denotemode{\dia\alpha\phi}a\in\scott\univ\s\lawson\univ$ and it suffices
   to show that $\denotemode{\dia\alpha\phi}a$ is
$\lawson\univ$-closed, i.e.\ $\scott\univ$-compact as an upper
set. By induction,
  $\denotemode\phi a$ is $\lawson\univ$-clopen; it is also
  $\scott\univ$-open so $\denotemode\phi a = \upp
   {F_\phi}{}$ for
    a finite subset $F_\phi\s\ce\univ$ as $\univ$ is algebraic.
    By the definition of $\denotemode{\dia\alpha\phi}a$,
    we have $e\in\denotemode{\dia\alpha\phi}a$
    iff $\modeup ea_\alpha\cap \upp {F_\phi}{}\not=\e$ iff $\modeup ea_\alpha\cap F_\phi\not=\e$
    (as $\modeup ea_\alpha$ is a lower set). For each $y\in F_\phi$
    define
    $c(y) = (c(y)_\gamma)_{\gamma\in\act}\in \univ$ by $c(y)_\beta\defeq (\e,\univ)$ for all
    $\beta\not=\alpha$; and $c(y)_\alpha\defeq (\down y{},\univ)$.
    Then $C\defeq \{c(y)\mid y\in F_\phi\}$ is finite and $C\s\ce \univ$.
    Since $y\in\modeup {c(y)}a_\alpha\cap F_\phi$ for all $y\in C$, we
    get
    $\upp C{}\s\denotemode{\dia\alpha\phi}a$ as the latter set is upper. Note that for each
    $y\in F_\phi$ we have
    $c(y)\leq e$ in $\univ$ iff $y\in\modeup ea_\alpha$.
    Therefore, $e\in\denotemode{\dia\alpha\phi}a$ implies
    $e\in\upp C{}$. Thus, $\denotemode{\dia\alpha\phi}a$
    equals $\upp C{}$ for the finite subset $C$ of
    $\ce \univ$.

  \item Let ${\it m} = {\it c}$. From Theorem~4.2 in~\cite{hjs04} we already
  know that $\denotemode{\dia\alpha\phi}c$ is $\scott\univ$-closed and therefore
  $\lawson\univ$-closed. Thus, it suffices to show that
  $\denotemode{\dia\alpha\phi}c$ is $\lawson\univ$-open.
  By induction, $\denotemode{\phi}c$ is $\lawson\univ$-open and
  therefore
    $D\setminus \denotemode{\phi}c =
    \denotemode{\neg\phi}a$ is $\lawson\univ$-closed (and
    $\scott\univ$-open), i.e.\ $\scott\univ$-compact upper.
    Since $\univ$ is algebraic, $\denotemode{\neg\phi}a = \upp {F_{\neg\phi}}{}$
    for a
    finite subset $F_{\neg\phi}$ of $\ce \univ$. Thus,
    $\denotemode{\phi}c = \univ\setminus \upp {F_{\neg\phi}}{}$. Inspecting the definition of
    $\denotemode{\dia\alpha\phi}c$,
     we infer $e\in\denotemode{\dia\alpha\phi}
    c$ iff there is some $x\in\modeup ec_\alpha$ such that
    $x\not\in\upp {F_{\neg\phi}}{}$. Now let $d\in\denotemode{\dia\alpha\phi}c$.
    We claim that there are compact elements $k$ and $l$ with
    $d\in \upp k{}\setminus \upp l{}\s\denotemode{\dia\alpha\phi}c$,
    which concludes the proof since $\upp k{}\setminus \upp l{}$ is
    $\lawson\univ$-open. Choose any $k\in \down d{}\cap\ce \univ$. As for
    $l = (l_\gamma)_{\gamma\in\act}$,
    set $l_\beta\defeq (\e,\univ)$ for all $\beta\not=\alpha$; and
    $l_\alpha\defeq (\e,\upp {F_{\neg\phi}}{})$; in particular, $l\in\ce \univ$.
    Note that $l\not\leq e$ in $\univ$
    iff $\modeup ec_\alpha\not\s \upp {F_{\neg\phi}}{}$ iff (for some $x\in \modeup ec_\alpha$,
    $x\not\in \upp {F_{\neg\phi}}{}$). Therefore, $d\in \upp k{}\setminus
    \upp l{}\s \denotemode{\dia\alpha\phi}c$.

  \end{itemize}
So $C_\Phi$ is $\lawson\univ$-closed as the intersection of
$\lawson\univ$-closed sets. \qed

\noindent In \cite{vickers89} open sets are thought of as
observable properties, so the denotations of Hennessy-Milner logic
formulas in $\univ$ (and in $\univx$) are closed under negation as
observations. If we extend these denotations to the modal
mu-calculus \cite{kozen83a}, we expect observable properties to
correspond to sets in the Borel algebra generated by
$\scott\univ$.

Using the denseness of embeddings of image-finite labelled
transition systems in $\univx$, we can prove the inclusion $\max
(\univx)\s C_\Phi$.

\begin{lem}\label{lemma:minc}
The set $\max (\univ)$ is contained in $C _\Phi$.
\end{lem}

\proof Let $A$ be the set of all embeddings $\embed {L,l}$ of
pointed image-\-finite labelled transition systems $(L,l)$. Then
$A\s C_\Phi$ follows as

\begin{itemize}
\item $(\mts D, \embed {L,l})$ is refinement-equivalent to
$(L,l)$, \item $\alpha\dia\phi\lor \neg\alpha\dia\phi$ is valid
over labelled transition systems for all $\phi$ of Hennessy-Milner
logic, \item $\bx{\delta _i}\phi$ is valid over labelled
transition systems whenever $\phi$ is, and \item $\modeup\models
a$ is the standard semantics of Hennessy-Milner logic over
labelled transition systems.
\end{itemize}

By Proposition~\ref{prop:dense}, $A$ is a dense subset of
$(\univx, \tau _\univx)$ and so its superset $C_\Phi\cap \max
(\univ)$ is also dense in $(\univx, \tau _\univx)$ and is $\tau
_\univx$-closed by the Lawson condition for $\univ$ since $C
_\Phi$ is $\lawson\univ$-closed by Lemma~\ref{lemma:denoteclosed}.
But the only dense $\tau _\univx$-closed subset of $(\univx, \tau
_\univx)$ is $\univx$ itself and so $C_\Phi\cap \max (\univ) =
\max (\univ)$ follows which implies $\max (\univ)\s C _\Phi$.\qed

For a proof of the reverse inclusion $C_\Phi\s \max (\univx)$ we
need to clarify the structure of elements in $C_\Phi$.

\begin{lem}\label{lemma:aboutc}
Let $d\in C _\Phi$. Then:

\begin{enumerate}
\item All $d'\in\univ$ that are reachable from $d$ in the labelled
transition system $(\univ, \modeup \Real c)$ are in $C_\Phi$ as
well.

\item For all $\alpha\in\act$ we have $\uppart d\alpha = \upp
{(\lowpart d\alpha\cap\uppart d\alpha)}{}$.

\item For all $\phi$ of Hennessy-Milner logic, $(\mts
D,d)\modeup\models c \phi$ implies $(\mts D,d)\modeup\models
a\phi$.

\end{enumerate}
\end{lem}

\proof\hfill 
\begin{enumerate}
\item Let $d'$ be reachable from $d$ in $(\univ,\modeup \Real c)$
and let $w'\in\act ^*$ be the word obtained by travelling from $d$
to $d'$ on such a path. Given $\formula{w}\alpha p\in \Phi$, the
concatenation $w'w$ is in $\act ^*$ and so $\formula {w'w}\alpha
p\in\Phi$. Thus the path for $w'$ above and $d\in C_\Phi$ ensure
$(\mts D,d')\modeup\models a\formula {w}\alpha p$ and so $d'\in
C_\Phi$.

\item Let $\alpha\in\act$. Since $\upp {(\lowpart
d\alpha\cap\uppart d\alpha)}{}\s \upp {\uppart d\alpha}{} =
\uppart d\alpha$, it suffices to show $\uppart d\alpha \s \upp
{(\lowpart d\alpha\cap\uppart d\alpha)}{}$. Proof by
contradiction: Let $x\in \uppart d\alpha\setminus \upp {(\lowpart
d\alpha\cap\uppart d\alpha)}{}$. Then $x\in \uppart d\alpha$ and
$\lowpart d\alpha\cap\uppart d\alpha\s \upp {(\lowpart
d\alpha\cap\uppart d\alpha)}{}$ imply $x\not\in \lowpart d\alpha$
and so $x\in\univ\setminus \lowpart d\alpha$. As $\univ$ is
algebraic and $\univ\setminus \lowpart d\alpha\in\scott\univ$,
there is some $\denoteD p\in\ce \univ$ with $\denoteD p\in
\univ\setminus \lowpart d\alpha$ and $\denoteD p\leq x$ and so
$\upp {\denoteD p}{}\cap \lowpart d\alpha = \e$ as $\lowpart
d\alpha$ is a lower set. But $d\in C_\Phi$ implies $(\mts
D,d)\modeup\models a \dia\alpha \phi _p\lor \neg\dia\alpha\phi
_p$, as $\dia\alpha \phi _p\lor \neg\dia\alpha\phi _p$ is
$\formula {\epsilon}\alpha p$, and so $\upp {\denoteD p}{}\cap
\lowpart d\alpha = \e$ implies $\upp {\denoteD p}{}\cap \uppart
d\alpha = \e$ by the definition of $\denotemode{\dia\alpha\phi}m$
in~(\ref{equ:diauniv}), contradicting $x\in \upp {\denoteD
p}{}\cap \uppart d\alpha$.

\item We use structural induction on $\phi$. The cases for
$\true$, negation, and conjunction are straightforward. Let $(\mts
D,d)\modeup\models c\dia\alpha\phi$, so $(\mts D,d')\modeup\models
c\phi$ for some $d'\in\uppart d\alpha$. By item~(2), there is some
$d''\in\lowpart d\alpha\cap \uppart d\alpha$ with $d''\leq d'$.
But then $(\mts D,d')\modeup\models c\phi$ and $d''\leq d'$ imply
$(\mts D,d'')\modeup\models c\phi$ by Corollary~\ref{cor:hjs04}.
Since $d''\in\uppart d\alpha$ is reachable from $d$ in
$(\univ,\modeup \Real c)$ it is in $C_\Phi$ by item~(1). Thus, we
can apply induction on $d''$ and get $(\mts D,d'')\modeup\models
a\phi$. Since $d''\in\lowpart d\alpha$, this renders $(\mts
D,d)\modeup\models a\dia\alpha\phi$.\qed

\end{enumerate}

We have now all the machinery at our disposal for stating and
proving our main results in the next two theorems.

\begin{thm}\label{theorem:one}
The set $\max (\univ)$ equals $C_\Phi$. In particular, $\max
(\univ)$ is $\lawson\univ$-closed and $(\univx,\tau _\univx)$ is a
Stone space in which the set of embeddings of pointed
image-\-finite labelled transition systems is dense.
\end{thm}

\proof From item~(3) of Lemma~\ref{lemma:aboutc} and
Lemma~\ref{lemma:aremax} we infer $C_\Phi\s \max (\univ)$.
Lemma~\ref{lemma:minc} then renders $\max (\univ) = C_\Phi$. By
Lemma~\ref{lemma:denoteclosed}, this means that $\max (\univ)$ is
$\lawson\univ$-closed. By Propositions~\ref{prop:zero}
and~\ref{prop:dense}, it suffices to show that $(\univx,\tau
_\univx)$ is compact. Let $\univx = \bigcup {\mathcal U}$ for
${\mathcal U}\s \tau _\univx$. By the definition of $\tau
_\univx$, each $U\in {\mathcal U}$ is of the form $V_U\cap \max
(\univ)$ for some $V_U\in\scott\univ$. Since $\univ$ is a bifinite
domain, $(\univ,\lawson\univ)$ is compact \cite{abramsky94}. Since
$\max (\univ)$ is $\lawson\univ$-closed it is
$\lawson\univ$-compact as a $\lawson\univ$-closed subset of the
compact space $(\univ,\lawson\univ)$. From $\univx = \bigcup
{\mathcal U}$ and $\scott\univ\s\lawson\univ$ we infer that $\max
(\univ)\s \bigcup \{V_U\mid U \in {\mathcal U}\}\s\lawson\univ$.
The $\lawson\univ$-compactness of $\max(\univ)$ therefore implies
the existence of a finite set ${\mathcal F}\s {\mathcal U}$ with
$\max (\univ)\s \bigcup \{V_U\mid U \in {\mathcal F}\}$. But then
$\univx\s \bigcup {\mathcal F}$ follows.\qed

\subsection{Maximal-points space as quotient space of labelled
transition systems} Theorem~\ref{theorem:one} is of interest in
its own right since $\max (D)$ is not $\lawson D$-closed for
bifinite domains $D$ in general. But we also have to demonstrate
that $\univx$ is the desired quotient space of labelled transition
systems modulo bisimulation.

\begin{defi}
Given a topological space $(X,\tau)$ let $\compact X\tau$ be the
\emph{poset of all $\tau$-compact subsets of $X$}, ordered by
reverse inclusion: $C\sqsubseteq C'$ iff $C'\s C$.
\end{defi}

\begin{thm}\label{theorem:two}\hfill 
\begin{enumerate}
\item The embedding $(M,i)\mapsto \embed {M,i}$ for pointed
image-\-finite modal transition systems given in \cite{hjs04}
extends to pointed modal transition systems such that labelled
transition systems are embedded into $\max(\univ)$.

\item Conversely, for any $d\in\max(\univ)$ the pointed mixed
transition system $(\mts D,d)$ is refinement-equivalent to a
labelled transition system. (It doesn't ``type check'' to ask
whether $(\mts D, d)$ is \emph{bisimilar} to a labelled transition
system; but $\down {}{}$ and $\upp {}{}$ are merely saturation
artifacts of the model.)

\item We have the isomorphism

\begin{equation}
\univx = \prod _{\alpha\in\act} \compact \univx{\tau _\univx}
\end{equation}

\noindent of sets where $x = (x_\alpha)_{\alpha\in\act}$ models
the $\alpha$-successors of $x$ as the $\tau _\univx$-compact set
$x_\alpha$, for each $\alpha\in\act$.

\end{enumerate}

\end{thm}

\proof\hfill 
\begin{enumerate}
\item Whenever a state $s$ has infinitely many states $\{s_i\mid
i\in I\}$ as $\alpha$-successors for $\modeup Rc$, choose a finite
subset $F$ of $I$, retain transitions $(s,\alpha,s_i)$ and their
must/may status for all $i\in F$, discard all $(s,\alpha,s_i)$
with $i\not\in F$, and create a \emph{may-stub} $s_F$ ($\denoteD
{s_F} = \bot _\univ$) and a may-transition $(s,\alpha,s_F)$. Doing
this for all events while, at the same time, unfolding $(M,i)$ as
a tree ensures that all approximations are image-\-finite with
limit $\embed {M,i}$ such that $(\mts D, \embed {M,i})$ is
refinement-equivalent to $(M,i)$. In particular, $\embed
{M,i}\in\max (\univ)$ by Lemma~\ref{lemma:aremax} whenever $(M,i)$
is a labelled transition system.

\item Let $d\in\max (\univ)$ and $\alpha\in\act$. The set
$\lowpart d\alpha\cap\uppart d\alpha$ is in $C _\Phi$, which
equals $\max (\univ)$, and $\uppart d\alpha = \upp {(\lowpart
d\alpha\cap\uppart d\alpha)}{}$ by Lemma~\ref{lemma:aboutc} and
Theorem~\ref{theorem:one}. Combining this with~(\ref{equ:mix}), we
infer $d = ((\down {(\lowpart d\alpha\cap\uppart d\alpha)}{},
\lowpart d\alpha\cap\uppart d\alpha))_{\alpha\in\act}$. But since
$C_\Phi$ is closed under states reachable in $(\univ, \modeup
\Real c)$, we may assume this representation for all elements $e$
reachable from $d$ in $(\univ, \modeup \Real c)$. Therefore,
$(\mts D, d)$ is refinement-equivalent to the modal transition
system with no may-transitions that replaces $\down {(\lowpart
e\alpha\cap\uppart e\alpha)}{}$ with $\lowpart e\alpha\cap\uppart
e\alpha$ for all $\alpha\in\act$ and all $e$ reachable from $d$ in
$(\univ, \modeup \Real c)$.

\item The isomorphism follows from the equation for $\univ$ and
Lemmas~34.5 and~25 of \cite{alessi03}; the latter is stated for
${\it SFP^M}$-domains $D$, which are bifinite, but its proof only
requires that $\max (D)$ is $\lawson D$-closed.\qed

\end{enumerate}

An immediate consequence of these two main theorems is that sets
of implementations of modal transition systems are compact in the
quotient space modulo bisimulation.

\begin{cor}\label{cor:cl}
For each pointed modal transition system $(M,s)$, its set of
implementations is compact in the quotient space of labelled
transition systems modulo bisimulation.
\end{cor}

\proof The set of implementations of $(M,s)$ in $\univx$ is $\meu
{\embed {M,s}} = \upp {\embed {M,s}}{}\cap \max (\univ)$ which is
$\lawson\univ$-closed as the intersection of two
$\lawson\univ$-closed sets and so it is $\tau_\univx$-compact.\qed

\section{Applications of compactness}\label{section:applications}
We now discuss some of the consequences of the compactness of
$\tau_\univx$: a compactness theorem for Hennessy-Milner logic on
compact sets of implementations, an abstract interpretation of
compact sets of implementations as Scott-closed sets of modal
transition systems, and a robust consistency measure for modal
transition systems.

\subsection{A compactness theorem for sets of
implementations} Compactness of $(\univx,\tau_\univx)$, stated in
terms of Hennessy-Milner logic, is familiar from first-order logic
but here secured without appeal to a complete proof system. Such
semantic techniques for proving compactness are not new, we
mention model-theoretic techniques based on ultra-filters. A
compactness theorem for Hennessy-Milner logic alone already
follows from its standard encoding in first-order logic. However,
we prove a compactness result that goes beyond Hennessy-Milner
logic as it applies to compact sets of labelled transition
systems, in particular to the set of common implementations of
finitely-many pointed modal transition systems. For a single such
system, $(\mts D,\bot _\univ)$, we then regain the familiar
compactness theorem for Hennessy-Milner logic. Our result is
stronger than this familiar theorem as the sets of implementations
of pointed modal transition systems are not expressible through
Hennessy-Milner logic. In Theorem~\ref{theorem:mc}(2) below we see
that these sets are expressible in Hennessy-Milner logic extended
with greatest fixed points for finite-state modal transition
systems.

\begin{cor}\label{coro:one}\hfill 
\begin{enumerate}
\item Let $\Gamma$ be a set of formulas of Hennessy-Milner logic
and $C$ a $\tau_\univx$-compact set such that for all finite
subsets $\Delta$ of $\Gamma$ there is some $c_\Delta\in C$ that
satisfies $\bigwedge \Delta$. Then there is some $c_\Gamma\in C$
that satisfies all formulas of $\Gamma$.

\item In particular, let $\Gamma$ be a set of formulas of
Hennessy-Milner logic and $\{ (M_i,s_i)\mid 1\leq i\leq k\}$ a
finite set of pointed modal transition systems such that for all
finite subsets $\Delta$ of $\Gamma$ there is a pointed labelled
transition system that refines all $(M_i,s_i)$ and satisfies
$\bigwedge \Delta$. Then there is a pointed labelled transition
system that refines all $(M_i,s_i)$ and satisfies all formulas of
$\Gamma$.
\end{enumerate}
\end{cor}

\proof By Corollary~\ref{cor:cl} it suffices to prove item~(1). By
duality of consistency (i.e.\ satisfiability) and validity, it
suffices to prove the dual statement of item~(1): assume that
every $c\in C$ satisfies as least one $\phi\in\Gamma$ and show
that there is a finite set $\Delta\s \Gamma$ such that $\bigvee
\Delta$ is valid over the set $C$. By this assumption, we have

\begin{equation}
C\s \bigcup {\mathcal U}
\end{equation}

\noindent where ${\mathcal U} = \{ \denotemode \phi a\cap \max
(\univ)\mid \phi\in \Gamma\}$ is a subset of $\tau_\univx$ as all
$\denotemode \phi a$ are in $\scott\univ$ by Theorem~4.2
in~\cite{hjs04}. As $C$ is $\tau_\univx$-compact, there is a
finite ${\mathcal F}\s {\mathcal U}$ with $C\s \bigcup {\mathcal
F}$, i.e.\ $C\s \bigcup _{\phi\in \Delta} \denotemode \phi a =
\denotemode {\bigvee \Delta} a$ for a finite set $\Delta\s
\Gamma$. Thus all $c\in C$ satisfy $\bigvee \Delta$.\qed

\begin{exa}
Figure~\ref{fig:cr} depicts schematically the set of common
implementations of two pointed modal transition systems $(\mts
D,d)$ and $(\mts D,e)$, the intersection of the implementations of
$d$ and $e$. This is a compact subset of $\univx$ and so we get a
compactness theorem for Hennessy-Milner logic on that set.
\end{exa}

\subsection{Abstract interpretation of $\tau_\univx$-compact sets of implementations}
Cousot \& Cousot's abstract interpretation framework
\cite{cousot77} approximates concrete objects and their
transformations by abstract objects and transformations such that
reasoning on abstract objects is sound for their concretizations.
In a simple setting, one has given a set $C$ of concrete objects
(e.g.\ computer programs) and a partial order $(A,\leq)$ of
abstract objects, a monotone abstraction function $\alpha\colon
(\ps C,\s)\to (A,\leq)$, and a monotone concretization function
$\gamma\colon (A,\leq)\to (\ps C,\s)$. The value $a = \alpha(X)$
should represent the best approximation of $X\s C$ within the
partial order $(A,\leq)$ and $\gamma(a)$ should represent the set
of those concrete objects that are abstracted by $a$. One can
encode these intuitions by making $\alpha$ and $\gamma$ a Galois
connection \cite{cousot77}, a notion we define below.

\begin{exa}
Let $C$ be the set of natural numbers and $A = \{\top, O,E\}$
where $\top$ is the top element and $O$ and $E$ are incomparable.
Define $\alpha (X)$ to be $O$ if all elements of $X$ are odd; $E$
if all elements of $X$ are even; and $\top$ otherwise. Then
$\alpha (\{2,46,128\}) = O$ and $\alpha (\{ 2,4,7\}) = \top$ etc.
Define $\gamma (E) = \{ 0,2,4,\dots\}$, $\gamma (O) =
\{1,3,5,\dots\}$, and $\gamma (\top) = C$. Then $\alpha (
\{2,46,128\}) = O$ says that $O$ is the least element that soundly
represents the set $\{2,46,128\}$. The equation $\gamma (\alpha
(\{2,46,128\})) = \{0,2,4,\dots\}$ shows that the abstract value
of $\{2,46,128\}$ has a larger set of concrete objects.
\end{exa}

We want to apply this framework in our setting. From the
compactness of $\tau_\univx$ Corollary~\ref{cor:cl} infers that
the set $\meu {\embed{M,s}}$ is $\tau_\univx$-compact for all
pointed modal transition systems $(M,s)$. Said
$\tau_\univx$-compact set comprises all the implementations of
$(M,i)$. Conversely, a $\tau_\univx$-compact set $C$ of labelled
transition system can be approximated by any pointed modal
transition system $(M,s)$ satisfying $C\s \meu {\embed {M,s}}$.
Ideally, one wants an \emph{optimal} such $(M,s)$, one for which
the difference $\meu {\embed {M,s}}\setminus C$ is minimal. Of
course, this optimality is ensured for any $C$ of the form $\meu
{\embed {M,s}}$. The next example shows that there is no optimal
$(M,s)$ in general.

\begin{exa}
Consider two pointed modal transition trees $(M_1,s_1)$ and
$(M_2,s_2)$ that have a common refinement but do not refine each
other. In general, there will be more than one minimal upper bound
of the set $\{\embed {M_1,s_1},\embed {M_2,s_2}\}$ in $\univ$ so
there cannot be a $d\in\univ$ such that $\meu d$ equals the
$\tau_\univx$-compact set $\meu {\embed {M_1,s_1}}\cap \meu
{\embed {M_2,s_2}}$.
\end{exa}

 The fact that modal transition systems cannot
be such optimal abstractions of $\tau_\univx$-compact sets seems
to be related to the incompleteness of modal transition systems
for abstraction-based model checking \cite{dams04} since $\univ$
is not bounded complete. But there is a Galois connection between
$\tau_\univx$-compact subsets of $\univx$ and $\scott\univ$-closed
subsets of $\univ$.  For a $\tau_\univx$-compact set $C$ its set
of concretizations is the Scott-closed set of all $(M,s)$ for
which $C\s \meu {\embed {M,s}}$. Conversely, a Scott-closed subset
$L$ of pointed modal transition systems is abstracted as the set
of those pointed labelled transition systems that implement all
elements of $L$.

\begin{defi}\hfill 
\begin{enumerate}
\item Let $\lpd \univ = \{L\mid L\ \mbox{$\scott\univ$-closed}\}$
be the set of $\scott\univ$-closed subsets of $\univ$, ordered by
set inclusion: $L$ is less than or equal to $L'$ iff $L\s L'$.

\item Let $L_1$ and $L_2$ be complete lattices. A \emph{Galois
connection} \cite{gierz80} is a pair of monotone maps
$\alpha\colon L_1\to L_2$ and $\gamma\colon L_2\to L_1$ such that
for all $x\in L_1$ we have $\gamma (\alpha(x))\geq x$ and for all
$y\in L_2$ we have $\alpha(\gamma(y))\leq y$. In that case,
$\alpha$ is the \emph{upper adjoint} of $\gamma$.
\end{enumerate}

\end{defi}

\begin{thm}\label{theorem:gc}
The maps $\gamma\colon \compact \univx{\tau_\univx}\to \lpd \univ$
and $\alpha\colon \lpd \univ\to \compact\univx{\tau_\univx}$
defined by

\begin{eqnarray}
\gamma (C) &=& \{ d\in\univ\mid C\s \meu d\}\\
\alpha (L) &=& \bigcap _{d\in L}\meu d\nonumber
\end{eqnarray}

\noindent form a Galois connection, where $\alpha$ is the upper
adjoint of $\gamma$.
\end{thm}

\proof\hfill 
 \begin{itemize} \item The map $\gamma$ is well defined.
First $d\leq e$ implies $\meu e\s\meu d$ and so $\gamma (C)$ is a
lower set. Second let $(d_i)_{i\in I}$ be directed in $\gamma
(C)$. Then $C\s \bigcap _{i\in I}\meu {d_i}$ and the latter equals
$\meu {\bigvee _{i\in I} d_i}$, so $\gamma (C)$ is
$\scott\univ$-closed.

\item The map $\alpha$ is well defined. For if $L$ is empty, then
$\alpha (L) = \univx$ is $\tau_\univx$-compact; and if $L$ is
non-empty, $\alpha (L)$ is the intersection of
$\lawson\univ$-closed elements and so $\lawson\univ$-closed whence
$\tau_\univx$-compact.

\item The map $\gamma$ is monotone. Let $C\sqsubseteq C'$, i.e.\
$C'\s C$. Then $d\in \gamma (C)$ means $C\s \meu d$ and so $C'\s
\meu d$ follows. Therefore $d\in\gamma (C')$ and so $\gamma (C)\s
\gamma (C')$.

\item The map $\alpha$ is monotone. Let $L\s L'$. Then $\alpha
(L') = \bigcap _{d\in L'}\meu d\s \bigcap _{d\in L}\meu d = \alpha
(L)$ and so $\alpha (L)\sqsubseteq \alpha (L')$.

\item To see $\gamma\circ\alpha\geq {\rm id}_{\lpd \univ}$ let
$L\in\lpd\univ$. Then $\gamma(\alpha(L)) = \{ e\in \univ\mid
\alpha(L)\s \meu e\} = \{ e\in \univ\mid \bigcap _{d\in L}\meu d\s
\meu e\}$ clearly contains $L$.

\item To see $\alpha\circ\gamma\leq {\rm id}_{\compact
\univx{\tau_\univx}}$ let $C\in\compact \univx{\tau_\univx}$. Then
$\alpha(\gamma(C)) = \alpha (\{ d\in\univ\mid C\s \meu d\}) =
\bigcap \{ \meu d\mid C\s \meu d\}$ obviously contains $C$.\qed

\end{itemize}

Theorem~\ref{theorem:gc} remains to be valid if we reverse the
orders on the domains $\compact \univx{\tau_\univx}$ and $\lpd
\univ$ and swap the names $\alpha$ and $\beta$ throughout the
theorem and its proof. In that case, a $\tau _{\univx}$-compact
set $C$ is abstracted by a set $L$ of pointed modal transition
systems and any such $L$ has a set of pointed labelled transition
systems as concretizations. This view is perhaps more natural.

\subsection{Consistency measure for modal transition systems}

We explicitly state the metrics $d_\univ$ for pointed modal
transition systems and $d_\univx$ for pointed labelled transition
systems. The latter is then used to define a consistency measure
on modal transition systems as an alternative to the metric
$d_\univ$. Fix an enumeration $p_0, p_1, \dots$ of $\mpa$ and set

\begin{eqnarray*}
d_{\univ} (d,e) &=& \inf \{ 2 ^{-n}\mid \forall i\leq
n\colon \denoteD {p_i}\leq d \hbox{ iff } \denoteD {p_i}\leq e \} \\
d_{\univx} (x,y) &=& \inf \{ 2 ^{-n}\mid \forall i\leq n\colon
\denoteD {p_i}\leq x \hbox{ iff } \denoteD {p_i}\leq y \} \,.
\end{eqnarray*}

Then the topology determined by $d_{\univ}$ and $d_{\univx}$ is
$\lawson\univ$ and $\tau _\univx$, respectively. For practical
purposes we wish to enumerate $p\in\mpa$ in increasing modal depth
of $\phi _p$ in~(\ref{equ:phip}), corresponding to the iterative
unfolding of the functional for bisimulation \cite{milner89}. In
that case, $d_\univx$ is essentially the metric in
\cite{BZ_acm82}. These metrics are standard and well understood
but result in consistency measures if lifted to compact sets of
implementations.

We define the \emph{consistency measure} $c = \lambda (d,e)\cdot
[c_1(d,e),c_2(d,e)]\colon \univ\times\univ\to\idom$ by

\begin{eqnarray*}
c_1(d,e) &=& \inf \{ d_{\univx}(x,y)\mid x\in\meu d,\ y\in\meu e\} \\
c_2(d,e) &=& \sup \{ d_{\univx}(x,y)\mid x\in\meu d,\ y\in\meu
e\}\nonumber
\end{eqnarray*}

\noindent and use this as an alternative to the metric $d_\univ$
for comparing the pointed modal transition systems $(\mts D,d)$
and $(\mts D,e)$. Note that $c_1$ and $c_2$ are optimistic and
pessimistic measures (respectively) from the point of view of an
implementor.

\begin{exa}
Figure~\ref{fig:cr} shows a scenario where two pointed modal
transition systems $(\mts D,d)$ and $(\mts D,e)$ have a common
refinement, and so $c_1(d,e) = 0$.
\end{exa}
\begin{figure}
\begin{center}
\setlength{\unitlength}{0.00052493in}
\begingroup\makeatletter\ifx\SetFigFont\undefined%
\gdef\SetFigFont#1#2#3#4#5{%
  \reset@font\fontsize{#1}{#2pt}%
  \fontfamily{#3}\fontseries{#4}\fontshape{#5}%
  \selectfont}%
\fi\endgroup%
{\renewcommand{\dashlinestretch}{30}
\begin{picture}(6222,3783)(0,-10)
\put(3240,3240){\blacken\ellipse{90}{90}}
\put(3240,3240){\ellipse{90}{90}}
\put(1620,3240){\blacken\ellipse{90}{90}}
\put(1620,3240){\ellipse{90}{90}}
\put(5040,3240){\blacken\ellipse{90}{90}}
\put(5040,3240){\ellipse{90}{90}}
\path(810,3240)(6210,3240)(3510,90)(810,3240)
\path(1575,3240)(2610,2115)(3780,3240)
\path(2565,3240)(4500,2475)(5040,3240)
\dashline{60.000}(1575,3465)(3735,3465)
\path(3735,3465)(3735,3375) \path(1575,3465)(1575,3375)
\path(4500,2475)(4501,2477)(4503,2481)
    (4507,2489)(4513,2501)(4521,2515)
    (4530,2533)(4540,2553)(4551,2575)
    (4562,2597)(4572,2621)(4583,2644)
    (4594,2668)(4603,2692)(4613,2717)
    (4621,2742)(4629,2766)(4635,2790)
    (4638,2807)(4641,2822)(4642,2835)
    (4642,2844)(4642,2851)(4641,2857)
    (4640,2860)(4639,2861)(4638,2862)
    (4637,2862)(4635,2861)(4633,2860)
    (4631,2860)(4629,2859)(4627,2859)
    (4624,2860)(4620,2861)(4616,2864)
    (4611,2867)(4605,2871)(4598,2876)
    (4590,2880)(4571,2887)(4555,2891)
    (4544,2892)(4536,2892)(4530,2891)
    (4524,2890)(4515,2888)(4501,2886)
    (4481,2884)(4455,2880)(4429,2875)
    (4409,2869)(4396,2864)(4389,2859)
    (4384,2854)(4379,2849)(4372,2844)
    (4360,2840)(4342,2837)(4320,2835)
    (4305,2836)(4292,2837)(4282,2839)
    (4275,2840)(4270,2841)(4266,2842)
    (4264,2843)(4261,2843)(4259,2845)
    (4255,2848)(4250,2852)(4244,2859)
    (4237,2868)(4230,2880)(4223,2899)
    (4221,2916)(4223,2929)(4228,2939)
    (4234,2948)(4240,2956)(4244,2966)
    (4245,2979)(4241,2996)(4230,3015)
    (4218,3029)(4206,3039)(4196,3045)
    (4189,3049)(4185,3051)(4181,3053)
    (4177,3053)(4171,3054)(4161,3055)
    (4145,3057)(4123,3059)(4095,3060)
    (4069,3059)(4046,3057)(4026,3054)
    (4011,3051)(4001,3048)(3993,3045)
    (3986,3041)(3980,3038)(3972,3034)
    (3960,3031)(3945,3027)(3924,3023)
    (3899,3019)(3870,3015)(3838,3012)
    (3813,3009)(3796,3006)(3786,3003)
    (3782,3000)(3780,2996)(3778,2993)
    (3774,2992)(3764,2992)(3747,2995)
    (3722,3003)(3690,3015)(3665,3027)
    (3642,3041)(3622,3054)(3607,3066)
    (3595,3077)(3586,3087)(3579,3096)
    (3574,3105)(3568,3114)(3562,3123)
    (3553,3133)(3542,3144)(3527,3156)
    (3509,3169)(3488,3183)(3465,3195)
    (3441,3206)(3419,3213)(3402,3217)
    (3387,3218)(3375,3218)(3364,3215)
    (3355,3212)(3348,3208)(3341,3204)
    (3336,3200)(3333,3198)(3331,3196)(3330,3195)
\path(2610,2115)(2608,2117)(2604,2121)
    (2598,2127)(2589,2137)(2578,2149)
    (2566,2164)(2554,2180)(2542,2197)
    (2530,2215)(2521,2233)(2514,2253)
    (2509,2273)(2508,2295)(2512,2317)
    (2520,2340)(2534,2361)(2550,2380)
    (2567,2393)(2584,2403)(2600,2410)
    (2616,2413)(2630,2414)(2644,2415)
    (2658,2415)(2673,2417)(2689,2419)
    (2706,2424)(2726,2433)(2747,2444)
    (2769,2459)(2790,2475)(2806,2490)
    (2819,2503)(2829,2514)(2837,2523)
    (2843,2529)(2847,2533)(2850,2536)
    (2852,2538)(2854,2539)(2855,2540)
    (2855,2542)(2856,2544)(2856,2549)
    (2855,2556)(2853,2566)(2849,2578)
    (2843,2593)(2835,2610)(2822,2630)
    (2807,2648)(2791,2664)(2775,2677)
    (2760,2687)(2746,2695)(2732,2702)
    (2719,2707)(2705,2713)(2692,2719)
    (2678,2727)(2663,2736)(2649,2747)
    (2634,2761)(2621,2775)(2610,2790)
    (2603,2809)(2608,2820)(2622,2823)
    (2641,2821)(2663,2816)(2684,2811)
    (2702,2809)(2714,2811)(2714,2820)
    (2700,2835)(2685,2845)(2667,2856)
    (2648,2865)(2627,2874)(2607,2881)
    (2586,2888)(2566,2893)(2547,2898)
    (2527,2903)(2508,2907)(2489,2912)
    (2470,2917)(2452,2924)(2434,2931)
    (2417,2940)(2402,2949)(2391,2960)
    (2385,2970)(2386,2980)(2394,2989)
    (2407,2994)(2423,2998)(2439,2999)
    (2457,2999)(2475,2998)(2494,2996)
    (2513,2995)(2532,2993)(2552,2993)
    (2573,2994)(2594,2996)(2616,3001)
    (2636,3007)(2655,3015)(2669,3024)
    (2678,3032)(2681,3038)(2678,3043)
    (2670,3047)(2660,3050)(2647,3051)
    (2633,3052)(2619,3054)(2607,3056)
    (2598,3059)(2594,3064)(2596,3071)
    (2606,3080)(2626,3092)(2655,3105)
    (2681,3115)(2711,3126)(2742,3136)
    (2774,3145)(2807,3154)(2840,3163)
    (2874,3171)(2908,3179)(2942,3187)
    (2976,3195)(3009,3202)(3042,3209)
    (3073,3215)(3102,3221)(3127,3227)
    (3149,3231)(3167,3234)(3179,3237)
    (3188,3239)(3193,3240)(3195,3240)
\put(855,1170){\makebox(0,0)[lb]{\smash{{{\SetFigFont{9}{10.8}{\rmdefault}{\mddefault}{\updefault}$c_1(d,e)
= 0$ as}}}}}
\put(3735,0){\makebox(0,0)[lb]{\smash{{{\SetFigFont{9}{10.8}{\rmdefault}{\mddefault}{\updefault}$\bot
_{\univ}$}}}}}
\put(4365,2250){\makebox(0,0)[lb]{\smash{{{\SetFigFont{9}{10.8}{\rmdefault}{\mddefault}{\updefault}$e$}}}}}
\put(0,810){\makebox(0,0)[lb]{\smash{{{\SetFigFont{9}{10.8}{\rmdefault}{\mddefault}{\updefault}$y$
is common refinement}}}}}
\put(2520,1890){\makebox(0,0)[lb]{\smash{{{\SetFigFont{9}{10.8}{\rmdefault}{\mddefault}{\updefault}$d$}}}}}
\put(3285,3060){\makebox(0,0)[lb]{\smash{{{\SetFigFont{9}{10.8}{\rmdefault}{\mddefault}{\updefault}$y$}}}}}
\put(1350,3015){\makebox(0,0)[lb]{\smash{{{\SetFigFont{9}{10.8}{\rmdefault}{\mddefault}{\updefault}$x$}}}}}
\put(1710,3600){\makebox(0,0)[lb]{\smash{{{\SetFigFont{9}{10.8}{\rmdefault}{\mddefault}{\updefault}implementations
of $d$}}}}}
\put(4635,3060){\makebox(0,0)[lb]{\smash{{{\SetFigFont{9}{10.8}{\rmdefault}{\mddefault}{\updefault}$z$}}}}}
\put(4230,585){\makebox(0,0)[lb]{\smash{{{\SetFigFont{9}{10.8}{\rmdefault}{\mddefault}{\updefault}$c_2(d,e)
= d_{\univx}(x,z)$}}}}}
\end{picture}
}
\end{center}
\caption{Two pointed modal transition systems $(\mts D,d)$ and
$(\mts D,e)$ that have a common refinement.\label{fig:cr}}
\end{figure}
Since $\meu f$ is $\tau _\univx$-compact for all $f\in\univ$ by
Corollary~\ref{cor:cl}, $c_1(d,e)$ and $c_2(d,e)$ are the metric
analogue of symmetric $\forall\forall$ and $\exists\exists$ lifts
of relations from elements to subsets, here of $d _\univx$ to
$\tau _\univx$-compact subsets, respectively. The standard metric
$c(d,e)$ between compact subsets $\meu d$ and $\meu e$, the
Hausdorff distance, is the symmetric $\exists\forall$-lift of
$d_\univx$ to $\tau _\univx$-compact subsets and so

\begin{equation}
c_1(d,e)\leq c(d,e)\leq c_2(d,e)\, .
\end{equation}

Such consistency measures are of particular interest if $d$ and
$e$ represent different viewpoints
\cite{nuseibeh94,jacksonviews95,sommerville98} of the same system
such that the degree of consistency between these descriptions
needs to be explored.

We prove that $c_1$ is a robust measure in that its kernel
consists of those pairs of pointed modal transition systems that
have a common refinement.

\begin{thm}\label{theorem:mc}\hfill 
\begin{enumerate}
\item For all $d,e\in\univ$, we have $c_1(d,e) = 0$ iff $(\mts
D,d)$ and $(\mts D, e)$ have a common refinement.

\item Deciding whether two finite-state modal transition systems
have a common refinement is reducible to checking the
satisfiability of a modal mu-calculus formula with greatest fixed
points only.
\end{enumerate}
\end{thm}

\proof\hfill 
\begin{enumerate}
\item We use Theorems~\ref{theorem:one} and~\ref{theorem:two}
repeatedly. If $(\mts D,d)$ and $(\mts D, e)$ have a common
refinement, there is some $m\in\meu d\cap\meu e$ and so $c_1(d,e)
= 0$ as $d_\univx (m,m) = 0$. Conversely, let $c_1(d,e) = 0$. Then
for each $n\geq 0$ there are $m^d_n\in\meu d$ and $m^e_n\in\meu e$
with $d_\univx(m^d_n,m^e_n) < 1/n$. Since $(\univx, \tau _\univx)$
is compact, there is a convergent subsequence $(m^d_{n_j})_{j\geq
0}$ of $(m^d_n)_{n\geq 0}$ with limit $m^d$ and so $m^d\in \meu d$
as the latter is $\tau _\univx$-closed. Since
$d_\univx(m^d_{n_j},m^e_{n_j}) < 1/{n_j}$ for each $j\geq 0$, this
implies $\inf \{ d_\univx (m^d,m^e_{n_j})\mid j\geq 0\} = 0$ and
so $m^d$ is in all $\tau _\univx$-closed sets that contain $\{
m^e_{n_j}\mid j\geq 0\}$. Therefore, $m^d$ is in $\meu e$ and so
$(\mts D, m^d)$ is a common refinement of $(\mts D,d)$ and $(\mts
D, e)$.

\item If $(M,i)$ and $(N,j)$ are finite-state, we show that there
are formulas $X _{(M,i)}$ and $X _{(N,j)}$ of the modal
mu-calculus with greatest fixed points only such that the modal
mu-calculus formula $X _{(M,i)}\land X _{(N,j)}$ is satisfiable
over labelled transition systems iff $(M,i)$ and $(N,j)$ have a
common refinement. Larsen \& Thomsen implicitly define these
formulas in the system of recursive equations~(3) of
\cite{larsen89b} where, for each state $s$ in $M=(\Sigma, \modeup
Ra,\modeup Rc)$,

\begin{equation}
X_{(M,s)} = (\bigwedge _{(s,\alpha,s')\in\modeup Ra} \dia\alpha
X_{(M,s')})\land (\bigwedge _{\alpha\in\act}\bx\alpha (\bigvee
_{(s,\alpha,s')\in\modeup Rc} X_{(M,s')}))
\end{equation}

\noindent as a greatest fixed point. If $s$ has finitely many
reachable states in $M$, then $X_{(M,s)}$ is expressible in the
modal mu-calculus, using a ``calling context'' on the set of
states $t$ that are $\modeup Rc$-reachable from $s$ and static
scoping of the greatest fixed-point operators $\nu Z_t.\phi$. Now
for all pointed labelled transition systems $(L,l)$ we have
$(L,l)\modeup\models a X_{(M,s)}$ iff $(M,s)\refine {} (L,l)$
where we can use the proof of~(3) in \cite{larsen89b} which works
in our setting as conjunctions and disjunctions need not be
finite.\qed
\end{enumerate}

\begin{exa}
Let $M$ be the modal transition system from Figure~\ref{fig:ref}.
We write $X_{(M,{\rm Drinks})}$ as a formula of the modal
mu-calculus with greatest fixed points only. Let

\begin{eqnarray}
X_{(M,{\rm Dr})} &=& \nu Z_{\rm Dr}. \bx {\rm drinks} Z_{\rm Dr}
\land \bx {\rm talks} X^{\rm Dr}_{(M,{\rm Ta})}\land \bx
{\rm orders} X^{\rm Dr}_{(M,{\rm Wa})}\\
X^{\rm Dr}_{(M,{\rm Ta})} &=& \nu Z_{\rm Ta}. \bx {\rm drinks}
Z_{\rm Dr} \land \bx
{\rm orders} X^{\rm Dr\, Ta}_{(M,{\rm Wa})}\nonumber\\
X^{\rm Dr}_{(M,{\rm Wa})} &=& \nu Z_{\rm Wa}. \dia {\rm newPint}
Z_{\rm Dr} \land \dia {\rm newPint} X^{\rm Dr}_{(M,{\rm Ta})}\land
\bx {\rm
newPint}(Z_{\rm Dr}\lor X^{\rm Dr\, Wa}_{(M,{\rm Ta})})\nonumber\\
X^{\rm Dr\, Wa}_{(M,{\rm Ta})} &=& \nu Z_{\rm Ta}. \bx {\rm
drinks} Z_{\rm Dr} \land \bx
{\rm orders} Z_{\rm Wa}\nonumber\\
X^{\rm Dr\, Ta}_{(M,{\rm Wa})} &=& \nu Z_{\rm Wa}. \dia {\rm
newPint} Z_{\rm Dr} \land \dia {\rm newPint} Z_{\rm Ta}\land \bx
{\rm newPint}(Z_{\rm Dr}\lor Z_{\rm Ta})\nonumber
\end{eqnarray}

\noindent where the superscripts in $X_{(M,s)}$ record the
``calling context'' of the recursions.
\end{exa}

\noindent So $c_1(d,e)$ measures the \emph{degree of
inconsistency} of $(\mts D, d)$ and $(\mts D, e)$, a lower bound
on the difference between their implementations, $c_2(d,e)$ is an
upper bound on such a difference, and none of them is a metric:
From item~(\ref{item:fourd}) of Definition~\ref{def:ultra}, $c_1$
satisfies only~(b) and $c_2$ satisfies only~(b) and~(c). The
reducibility of common refinement checks to satisfiability checks
in the modal mu-calculus yields EXPTIME as a weak upper bound on
its complexity. Since the formulas are defined in terms of
greatest fixed points only, one can indeed show a stronger result:
the decision problem of common refinements is in PTIME
\cite{huth_isola04}.

\subsection{Scope of these results}
Our results also apply to 3-valued model checking frameworks in
which system observables are state propositions or a combination
of state propositions and events. This is so since Godefroid \&
Jagadeesan's translation between modal transition systems (events
only), partial Kripke structures \cite{bruns99} (state
propositions only), and Kripke modal transition systems
\cite{hjs01} (events and state propositions) and their
translations of the respective temporal logic formulas is shown to
preserve and reflect refinement and the meaning of model checks
\cite{gj03}.

\section{Related work}\label{section:related} Bakker \& Zucker use
domain equations and metric completions for a metric and
denotational treatment of concurrency in \cite{BZ_acm82}.

Lawson proposes the notion of a maximal-point space to represent
classical topological spaces as maximal points of a domain in the
topology induced by the domain's Lawson- and Scott-topology
\cite{lawson97}.

Abramsky \cite{abramsky91} provides a fully abstract domain of
synchronization trees for \emph{partial} bisimulation between
labelled transition systems that have a divergence predicate. The
domain equation of loc.\ cit.\ uses a sum construction on the
convex powerdomain. Maximal points are not part of that paper's
agenda and are therefore not discussed therein. Labelled
transition systems with a divergence predicate and partial
bisimulation are recognized as certain modal transition systems
and their refinement in \cite{hjs01}.

Mislove et al.\ present a fully abstract domain model, which
combines the probabilistic power domain with a convex variant of
the Plotkin powerdomain, for finite-state processes with
non-deterministic and probabilistic choice \cite{mislove03}.

Alessi et al.\ \cite{alessi03} introduce a category of ${\it
SFP^M}$-domains with a compositional maximal-points space functor
to Stone spaces. They show that all bifinite domains $D$ for which
$\max (D)$ is a Stone space are Scott-continuous retracts of ${\it
SFP^M}$-domains. In particular, $\univ$ is such a retract by
Theorem~\ref{theorem:one}. We suspect that $\univ$ is not an ${\it
SFP^M}$-domain since $\mpd {D_1}$ is not an ${\it SFP^M}$-domain
for the ${\it SFP^M}$-domain $D_1 = \{ \bot < \false, \true\}$
\cite{alessi97}, although $\mpd {D_1}$ is the second iteration of
the domain equation~(\ref{equ:recursion}) for $\univ$ when $\act =
\{\alpha\}$.

The paper \cite{hjs04} presents the domain $\univ$ and its modal
transition system $\mts D$, both denoted as $\mts D$ in loc.\
cit., and proves full abstraction and a characterization of
$\univ$'s compact elements in terms of formulas of Hennessy-Milner
logic.

In \cite{huth_refinement04} it is shown that the co-inductive
refinement of modal transition systems has an extensional
description: a pointed modal transition system $(M,i)$ refines a
pointed modal transition system $(N,j)$ if, and only if, the set
of implementations of $(M,i)$ is a subset of the implementations
of $(N,j)$.

Dams \& Namjoshi \cite{dams04} show that finite-state modal
transition systems are incomplete as abstractions of
infinite-state modal transition systems for modal mu-calculus
checking. They propose focused transition systems as a
generalization of modal transition systems, show completeness for
this class of models, and define a game semantics for refinement
of focused transition systems and a game semantics for model
checks of alternating tree automata on focused transition systems.
It is straightforward to write down a domain equation for focused
transition systems but a programme of maximal-points spaces won't
directly render pointed Kripke structures since, as noted in
\cite{dams04}, focused transition systems can have maximal
refinements that have inconsistent constraints on propositions at
states.

In \cite{huth_isola04} consistency, satisfiability, and validity
problems are studied for collectively model checking a set of
views endowed with labelled transitions, hybrid constraints on
states, and atomic propositions. A PTIME algorithm for deciding
whether a set of views has a common refinement (consistency) is
given. It is proved that deciding whether a common refinement
satisfies a formula of the hybrid mu-calculus \cite{sattler01}
(satisfiability), and its dual (validity), are EXPTIME-complete.
Two generically generated ``summary'' views are defined that
constitute informative and consistent common refinements and
abstractions of a set of views (respectively).

Di~Pierro et al.\ \cite{dipierrojcs02} develop a quantitative
notion of process equivalence as the basis for an approximative
version of non-interference and precise quantifications of
information leakage. They present two semantics-based analyzes for
approximative non-interference where one soundly abstracts the
other.

Desharnais et al.\ \cite{radha00} show that each continuous-state
labelled Markov process has a sequence of finite acyclic labelled
Markov processes as abstractions which is precise for a
probabilistic modal logic; an equivalence between the category of
Markov processes and simulation morphisms and a recursively
defined domain, viewed as a category, is given.

Desharnais et al.\ \cite{radha02} define a pseudo metric between
labelled concurrent Markov chains where zero distance means weak
bisimilarity. The metric is characterized in a real-valued modal
logic and shown to allow for compositional quantitative reasoning.

\section{Conclusions}\label{section:conclude} We presented the
fully abstract and universal domain model $\univ$ for pointed
modal transition systems and refinement of \cite{hjs04}. Using
techniques from concurrency theory and topology, we demonstrated
that $\univ$ is the right fully abstract and universal model for
labelled transition systems and bisimulation since the quotient
space of all pointed labelled transition systems with respect to
bisimulation, $(\univx,\tau _\univx)$, is obtained as the
maximal-points space of $\univ$. We furthermore revealed the
fine-structure of $\univx$, notably we proved that its topology
$\tau _\univx$ inherited from the Scott- and Lawson-topology of
$\univ$ is compact, zero-dimensional, and Hausdorff (a Stone
space). In particular, $\tau _\univx$ is determined by a
computationally meaningful, complete ultra-metric $d_{\univx}$ for
which image-\-finite labelled transition systems approximate
labelled transition systems to any degree of precision. Modulo
refinement, $(\mts D, k)$ is image-finite for all $k\in\ce\univ$,
so this denseness also applies to modal transition systems for the
Lawson-topology and its metric $d_\univ$. Thus our results unify
denotational, operational, and metric semantics of labelled and
modal transition systems. We finally derived consequences of this
compact representation: a compactness theorem for Hennessy-Milner
logic on compact sets of implementations, an abstract
interpretation of compact sets of implementations as Scott-closed
sets of modal transition systems, and a robust consistency measure
for modal transition systems.

\section*{Acknowledgment} Radha Jagadeesan suggested working with
the mixed powerdomain in \cite{hjs04}. Glenn Bruns, Alessandra
Di~Pierro, Patrice Godefroid, Dimitar Guelev, Chris Hankin, Radha
Jagadeesan, Achim Jung, Ralph Kopperman, David Schmidt, and
Herbert Wiklicky are thanked for helpful comments and discussions.
This paper is an extended journal version of \cite{huth_lics04}
and reflects the thorough and thoughtful comments made by the
anonymous referees of the LICS 2004 conference and the journal
Logical Methods in Computer Science.

\end{document}